\documentclass[authoryear,preprint,review,12pt]{elsarticle}

\usepackage{amssymb}
 \usepackage{amsthm}
 \usepackage{amsmath}
 \usepackage{bm}
 \usepackage{mediabb}
 \usepackage[mathlines, displaymath]{lineno}
 \usepackage{color}

\biboptions{longnamesfirst,comma}

\journal{Icarus}

\begin{document}
\begin{frontmatter}



\title{Impact cratering mechanics: A forward approach to predicting ejecta velocity distribution and transient crater radii}


\author[CIT]{Kosuke Kurosawa
\footnote{Corresponding author\\Kosuke Kurosawa, Ph.D.\\Planetary Exploration Research Center, Chiba Institute of Technology\\
  E-mail: kosuke.kurosawa@perc.it-chiba.ac.jp\\
  Tel: +81-47-4782-0320\\
  Fax: +81-47-4782-0372}}
\author[ERI]{Satoshi Takada}
\address[CIT]{Planetary Exploration Research Center, Chiba Institute of Technology, 2-17-1, Narashino, Tsudanuma, Chiba 275-0016, Japan}
\address[ERI]{Earthquake Research Institute, The University of Tokyo, 1-1-1, Yayoi, Bunkyo-ku, Tokyo 113-0032, Japan}

\begin{abstract}
Impact craters are among the most prominent topographic features on planetary bodies. 
Crater scaling laws allow us to extract information about the impact histories on the host bodies. 
The $\pi$-group scaling laws \citep[e.g.,][]{Holsapple1982} have been constructed based on the point-source approximation, dimensional analysis, and the results from laboratory and numerical impact experiments. 
Recent laboratory and numerical impact experiments, however, demonstrated that the scaling parameters themselves exhibits complex behavior against the change in the impact conditions and target properties. 
Since impact experiments are expensive and time-consuming in terms of obtaining new scaling constants, it is not feasible to explore the entire parameter space via experiments.
Here, we propose an alternative, fully analytical method to predict impact outcomes, including the ejection velocity distribution and transient crater radii, based on impact cratering mechanics. 
This approach is based on the Maxwell $Z$-model [Maxwell, 1977, Impact and Explosion Cratering, New York: Pergamon Press, pp. 1003--1008] and the residual velocity [Melosh, 1985, {\it Icarus} \textbf{62}, 339--343]. 
Given that the shapes of the streamlines of the excavation flow and the kinetic energy in a given streamtube are known, we can calculate the ejecta velocity distribution and investigate the cessation of crater growth. 
We present analytical expressions of (1) the proportionality relation between the ejection velocity and the ejection position, (2) the radius of a growing crater as a function of time, and (3) the transient crater radii in the gravity- and strength-dominated regimes. 
Since we focused on obtaining analytical solutions in this study, a number of simplifications are employed, such as a priori assumption of the direction of the velocity vectors of the excavating materials, the neglect of the effects of dry friction, metal-like targets with a constant yield strength. 
Due to the simplifications in the strength model, the accuracy of the prediction in the strength-dominated cratering regime is relatively low. 
Our model reproduces the power-law behavior of the ejecta velocity distribution and the approximate time variation of a growing crater predicted by $\pi$-group scaling laws. 
In our model, the transient crater radius depends strongly on the shape exponent $Z$, the shock decay exponent $n$, and the exponent $m$ pertaining to the residual velocity. 
Thus, the nature of shock propagation and the thermodynamic response of the shocked media, which cannot be addressed by dimensional analyses as a matter of principle, are naturally included in our estimation. 
The predicted radii under typical impact conditions mostly converge to a region between the two typical scaling lines for dry and wet sands predicted by the $\pi$-group scaling laws, strongly supporting the notion that the new method is one of the simplest ways to predict impact outcomes, as it provides analytical solutions. 
Our model could serve as a quick-look tool to estimate the impact outcome under a given set of conditions, and it might provide new insights into the nature of impact excavation processes. (463 words)
\end{abstract}


\end{frontmatter}


\section{Introduction}
\label{sec:Introduction}
Impact craters are among the dominant geographical features on planets, satellites, and small bodies without a hydrosphere or atmosphere. 
Craters provide evidence that the host body has suffered intense impact bombardment throughout its history \citep[e.g.,][]{Neukum1994, Ryder2002, Robbins2014, Fassett2016}. 
The crater size and ejecta deposits around the host crater, as observed using remote sensing methods, could constrain the impact history on a given planetary body. 
Thus, the relationship between impact conditions and impact outcomes has been investigated extensively using both experimental and numerical methods. 
The widely used $\pi$-group scaling laws have been constructed using such information about crater formation \citep[e.g.,][]{Holsapple1982, Schmidt1987, Holsapple1993, Johnson2016, Prieur2017}.

The $\pi$-group scaling laws have been constructed based on the `point-source theory' and dimensional analysis \citep[e.g.,][]{Buckingham1914, Dienes1970, Holsapple1982, Holsapple1987}.
First, we briefly discuss the point-source theory. 
It is widely believed that impact-related processes during the late stages of impact phenomena, including the crater radius and the ejecta velocity distribution, can be described by a single quantity, the coupling parameter $C$ \citep[e.g.,][]{Dienes1970, Holsapple1982, Holsapple1993}, as follows:
\begin{equation}
	C = R_{\rm p} v_{\rm imp}^\mu \rho_{\rm p}^\nu,
\end{equation}
where $R_{\rm p}$, $v_{\rm imp}$, $\mu$, $\rho_{\rm p}$, and $\nu$ are the projectile radius, the impact velocity, a velocity-scaling exponent, the projectile density, and a density-scaling exponent, respectively. 
The presence of coupling parameter was originally reported as ``the late-stage equivalence'' based on a series of numerical experiments, which modeled collisions between two identical metals \citep{Dienes1970}. 
Subsequently, the term $\rho_{\rm p}^\nu$ was introduced to address a density contrast between the projectile and the target \citep[e.g.,][]{Holsapple1982}. 
The velocity-scaling exponent $\mu$ was estimated at $0.58\pm0.01$ for consolidated materials \citep{Dienes1970}. 
This value is applicable for impact velocities well above the target sound speed. 
The exponent $\mu$ can range from $1/3$ (momentum scaling) to $2/3$ (energy scaling) under the different impact conditions \citep[e.g.,][]{Dienes1970}.
\citet{Mizutani1983, Mizutani1990} have pointed out that the late-stage equivalence holds only for an intermediate range of shock pressure and that $\mu$ is related to the pressure decay exponent in the pressure range. 
In the late-stage equivalence, the point-source approximation is the most important assumption.

Second, we describe the concept of the dimensional analysis. 
Using seven variables related to the diameter of a transient crater $D_{\rm tr}$, the impact velocity $v_{\rm imp}$, the projectile diameter $D_{\rm p}$, gravitational acceleration $g$, the strength of the target body $Y$, projectile density $\rho_{\rm p}$, and target density $\rho_{\rm t}$, four independent dimensionless parameters ($\pi_D$, $\pi_2$,$\pi_3$, and $\pi_4$) can be derived: 
\begin{eqnarray}
	\pi_{\rm D}&=& D_{\rm tr}\left(\frac{\rho_{\rm t}}{M_{\rm p}}\right)^{\frac{1}{3}}, \label{eq:pi_D}\\
	\pi_2 &=& \left(\frac{4\pi}{3}\right)^{\frac{1}{3}} \frac{gD_{\rm p}}{v_{\rm imp}^2}
		= \frac{1}{16}\left(\frac{4\pi}{3}\right)^{\frac{4}{3}} \frac{\rho_{\rm p} gD_{\rm p}^4}{E_{\rm proj}}, \label{eq:pi_2}\\
	\pi_3 &=& \frac{Y}{\rho_{\rm t}v_{\rm imp}^2},\label{eq:pi_3}\\
	\pi_4 &=& \frac{\rho_{\rm t}}{\rho_{\rm p}},
\end{eqnarray}
where $M_{\rm p} = (\pi/6)\rho_{\rm p}D_{\rm p}^3$ and $E_{\rm proj} = M_{\rm p}v_{\rm imp}^2/2$ are the projectile mass and the initial kinetic energy of the impactor, respectively. 
The four parameters $\pi_D$, $\pi_2$,$\pi_3$, and $\pi_4$ are often referred to as the scaled crater diameter, the gravity-scaled size, the non-dimensional strength, and the density ratio, respectively. 
Detailed descriptions of all dimensionless variables can be found in the literature \citep[e.g.,][]{Melosh1989}. 
By combining these parameters with the coupling parameter, the functional relationship between a dimensionless measure of the crater diameter $\pi_D$ and the other variables is obtained as follows \citep{Holsapple1993}:
\begin{equation}
	\pi_{\rm D} = K_1 \left[ \pi_2 \pi_4^{-\frac{\mu+2-6\nu}{3\mu}}+ \left(\pi_3 \pi_4^{-\frac{2-6\nu}{3\mu}}\right)^{\frac{\mu+2}{2}}\right]^{\frac{-\mu}{\mu+2}}, \label{eq:pi_D_K1}
\end{equation}
where $K_1$ is a scaling constant. 
Note that Eq.\ (\ref{eq:pi_D_K1}) is a full description, which covers both the gravity-dominated ($\pi_2\gg \pi_3$) and strength-dominated ($\pi_2 \gg \pi_3$) regimes. 
A similar formulation pertaining to the ejecta velocity distribution has been proposed \citep[e.g.,][]{Housen1983, Housen2011}. 
It should be mentioned that, in principle, dimensional analysis does not provide absolute values, including the position-dependent ejection velocity and the transient crater radius. 
Thus, the scaling parameters, including $K_1$, $\mu$, and $\nu$, have been widely explored empirically based on both laboratory and numerical experiments \citep[e.g.,][]{Gault1973, Gault1977, Schmidt1980, Schmidt1987, OKeefe1993, Cintala1999, Wunnemann2006, Wunnemann2011, Wunnemann2016, Yamamoto2006, Baldwin2007, Elbeshausen2009, Kraus2011, Kenkmann2011, Suzuki2012, Guldemeister2015, Prieur2017}. 
However, existing data pertaining to crater radii have not converged to the single universal line predicted by Eq.\ (\ref{eq:pi_D_K1}) \citep[e.g.,][]{Melosh1989}. 
In addition, recent laboratory/numerical impact experiments show that the scaling parameters themselves exhibits a complex behavior against the change in the impact/target conditions, such as impact velocity $v_{\rm imp}$ \citep{Barnouin-Jha2007, Yamamoto2017}, internal friction $f$ and porosity $\phi$\citep[e.g.,][]{Wunnemann2006, Elbeshausen2009, Prieur2017}.

The complexity of the scaling laws, as discussed in the previous paragraph, might originate from the limitations pertaining to dimensional analysis, which in principle cannot describe the mechanical aspects of impact cratering processes. 
In addition, both laboratory and numerical experiments are highly time-consuming and expensive in terms of exploring the entire parameter space. 
Thus, the objective of this study is to develop the first fully analytic method to estimate crater size and the ejection velocity distribution, without reliance on the $\pi$-group scaling laws. 
Although the model requires several assumptions or simplifications, as discussed below, it can provide the full chain of relations connecting the projectile/target parameters with the impact outcomes, including the ejecta velocity distribution and the transient crater radii. 
Since the fully analytic model allows us to quickly examine the parameter dependence on the impact outcomes, it would greatly help to minimize the required number of shots in future laboratory and numerical experiments to obtain new, high-accuracy scaling parameters.
Here, we briefly describe the steps included in the model, as follows: (1) estimate at each target point the peak particle velocity at the shock front, (2) remap the peak-particle velocity to the prescribed incompressible flow field, (3) evaluate the kinetic energy in each streamtube, (4) calculate the residual kinetic energy at the moment when the materials in a streamtube will lift above the pre-impact target surface, and (5) find the streamtube where the initial kinetic energy would be completely spent on work as a function of gravity and strength energy. 
Step (4) allows us to estimate the ejecta velocity at the distance where the streamtube leaves the target, and step (5) allows us to determine the transient cavity radius. 

Two key models are used in steps (2) and (3): the concept of the residual velocity and Maxwell's $Z$-model, respectively. 
The former was proposed by \citet{Melosh1985b}, who derived a relationship between shock propagation and the subsequent excavation flow based on thermodynamics. 
Although the behavior of shock propagation has been studied extensively \cite[e.g.,][]{Perret1975}, the link between the initial compressible radial flow and a late-stage incompressible excavation flow was unknown at the time.
\citet{Melosh1985b} pointed out that the residual velocity $u_{\rm p,res}$, which corresponds to the particle velocity upon the arrival of a subsequent expansion wave, is the origin of the normal excavation flow. 
During a shock-release cycle, the absolute magnitude of the particle velocity is significantly reduced owing to the expansion toward the free surface, and the direction of the velocity vectors of the shocked materials is changed significantly from that of the downward-propagating shock wave. 
The shocked materials with residual velocities form an excavation flow directed upward. 
The magnitude of the residual velocity can be estimated from thermodynamics by employing an equation of state (EOS) and the integral of the Riemann invariant along the isentrope from the shocked to the reference state. 
The latter analytic model was constructed by \citet{Maxwell1977} to predict the geometry of the late-stage incompressible flow. A combination of the residual velocity and the streamlines calculated based on the $Z$-model allows us to accomplish steps (2) and (3). 


\section{Rationale}\label{sec:2}
In this section, we describe the basic principles of our model. 
The Maxwell $Z$-model is briefly described in Section \ref{sec:2.1}, and the procedure used to calculate the residual velocity following a shock-release cycle from the peak-particle-velocity distribution is described in Section \ref{sec:2.2}. 
Finally, in Section \ref{sec:2.3} we discuss how to calculate crater radii and the ejecta velocity distribution under a given impact condition.


\subsection{Shapes of the streamlines of the excavation flow}\label{sec:2.1}
\begin{figure}[htbp]
	\begin{center}
		\includegraphics[width=\linewidth]{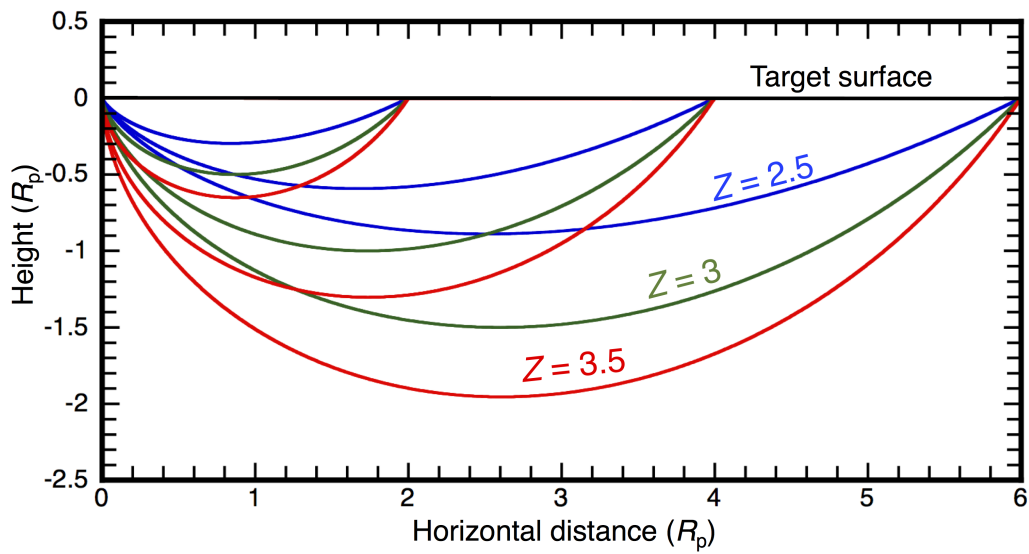}
	\end{center}
	\caption{Examples of the shapes of streamlines predicted by the Maxwell $Z$-model. 
	We used $Z = 2.5$ (blue), $3.0$ (green), and $3.5$ (red). 
	The $x$ and $y$ axes are normalized by the projectile radius, $R_{\rm p}$.}
	\label{fig:fig1}
\end{figure}

We use the $Z$-model to construct the geometry of the excavation flow. 
The radial velocity $u_r$ below the pre-impact surface is as follows \citep{Maxwell1977}:
\begin{equation}
	u_r = \alpha(t) r^{-Z}, \label{eq:u_r}
\end{equation}
where $\alpha(t)$, $r$, and $Z$ are the time-dependent strength of the excavation flow, the distance from the impact point, and a decay exponent that determines the curvature of the flow field, respectively. 
If the excavation flow is incompressible, the angular component of the flow velocity $u_\theta$ in polar coordinates $(r, \theta)$ can be calculated using the incompressibility constraint $\bm\nabla \cdot \bm{u}=0$. 
The geometry of the streamlines is given by
\begin{equation}
	r = R(1-\cos\theta)^{\frac{1}{Z-2}}, \label{eq:r_R_theta}
\end{equation}
where $R$ is the horizontal distance from the impact point to the intersection of a given streamline and the pre-impact surface. 
The $\theta=0\ $axis is directed vertically downward. 
Figure \ref{fig:fig1} shows examples of the shapes of streamlines considered here. 
To describe the impact excavation, $Z$ must exceed two. 
If $Z = 2$, the excavation flow becomes a purely radial flow, and it never reaches the surface for $Z \le 2$.
Note that $\alpha(t)$ is not needed to describe the shapes of the streamlines and that the explicit form of $\alpha(t)$ is not given by the $Z$-model on its own. 
Thus, one of the simplest assumptions (i.e., that $\alpha(t)$ is time-independent) has been used frequently \citep[e.g.,][]{Housen1983}. 
In other words, such previous studies used the $Z$-model only to describe the streamlines in a steady state. 
The cessation of crater growth, however, cannot be addressed by assuming $\alpha(t)={\rm Const.}$ 

\begin{figure}[htbp]
	\begin{center}
		\includegraphics[width=\linewidth]{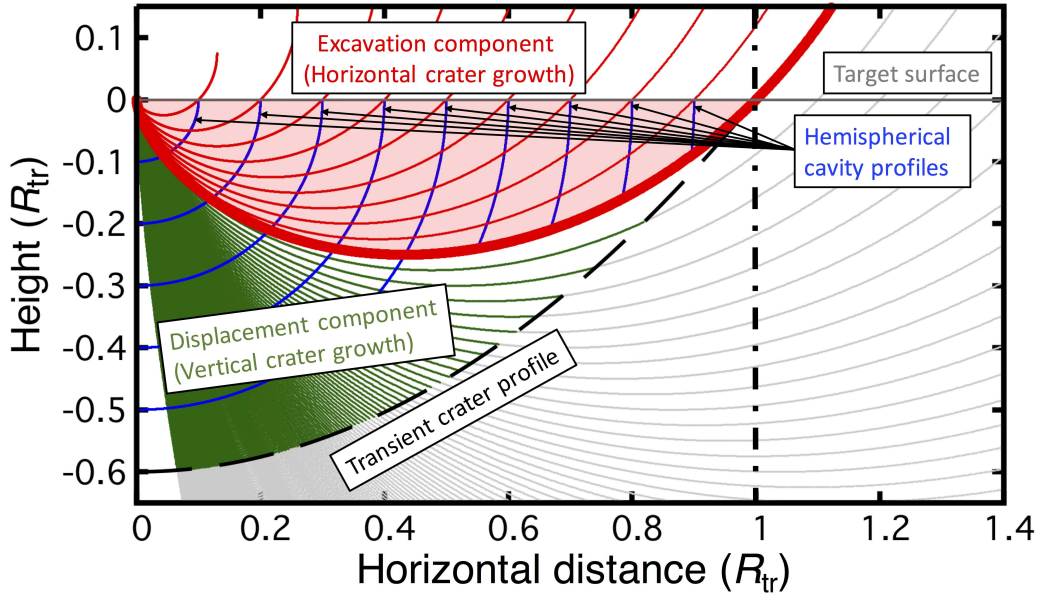}
	\end{center}
	\caption{Schematic diagram of the excavation and displacement components of the flow field in the $Z$-model (see Section \ref{sec:2.1}). 
	The horizontal and vertical axes are normalized by the transient crater radius $R_{\rm tr}$. 
	The red curves are for the excavation component. 
	The red shaded region has been excavated from the target. 
	The blue curves indicate the hemispherical transient-cavity profiles originally predicted by the $Z$-model. 
	The thick black dashed curve represents the actual cavity profile of a transient crater. 
	The green curves correspond to the flow in the displacement component. 
	We assumed that the hemispherical cavity growth proceeds approximately until $R_{\rm cr} = 0.5R_{\rm tr}$, where $R_{\rm cr}$ is the radius of a growing crater. 
	At some later time, the target's displaced volume reaches a maximum owing to depth-dependent effects (see Section \ref{sec:2.1}), while the materials in the excavation component continue to move, having been ejected above the target surface. }
	\label{fig:fig2}
\end{figure}

In this study, we only discuss crater growth in the horizontal direction. Here, we briefly discuss the difficulties in estimating the vertical growth, as follows. 
The original version of the Maxwell $Z$-model predicts that the shape of the growing cavity should be a hemisphere, because the radial component of the particle velocity $u_r$ exhibits a one-dimensional form as a function of distance, $r$ (Eq.\ (\ref{eq:u_r})). 
In contrast, it is known that the actual crater growth on most geologic materials, like sand and fragmented rocks, is characterized by two stages, as follows. 
Hemispheric cavity growth stops when its maximum depth is reached, and crater growth in the horizontal direction ceases on timescales that are several times longer \citep{Barnouin-Jha2007, Yamamoto2009, Yamamoto2017}. 
Hence, we divide the streamlines predicted by the $Z$-model into ``excavation'' and ``displacement'' components. 
The former corresponds to streamlines within the transient crater radius; i.e., the material covered by this component is ejected from the pre-impact target surface. 
The material associated with the latter component is never launched from the surface, although this part contributes to structural uplift during the final phase of crater formation. 
Thus, the former and latter components contribute mainly to crater growth in the horizontal and vertical directions, respectively. 
Figure \ref{fig:fig2} shows a schematic diagram of the components' $Z$-trajectories. 
In realistic cases, the earlier cessation of crater growth in the vertical direction is, as mentioned above, expected to result from resistance to displacement owing to dynamic rebound, which comes from the pressure gradient produced by the isostatic pressure, and the depth-dependent strength caused by dry friction in the geologic media through isostatic pressure. 
Thus, the actual crater growth will deviate from the hemispherical cavity growth predicted by the $Z$-model. 
Such dynamic effects cannot be treated based on an analytical approach without the functional form of the time-dependent flow strength, $\alpha(t)$. 
In contrast, the streamlines in the excavation component, which mainly contribute to growth in the horizontal direction, can be approximated by the $Z$-model, as they are not strongly affected by depth-dependent effects. 
Note that this treatment is unphysical in a strict sense, because a huge strain is produced between both components along the boundary.

There is another reason why we do not address the crater depth in this study. 
In reality, the central point of the excavation flow is located somewhat below the pre-impact surface \citep[e.g.,][]{Croft1980}.
Nevertheless, we decided to neglect this burial depth in the following discussion, as the effect of burial depth on crater radius is relatively small, although it significantly affects the crater depth \citep[e.g.,][]{Croft1980, Stewart2006, Kurosawa2015}. 
This simplification allows us to easily solve the system of equations analytically. 
Figure \ref{fig:fig1} shows examples of the shapes of streamlines considered here. 
To describe the impact excavation, $Z$ must exceed two. 
If $Z = 2$, the excavation flow becomes a purely radial flow, and it never reaches the surface for $Z < 2$.

\subsection{Residual velocity following a shock-release sequence}\label{sec:2.2}
Here, we describe the key concept of our model, based on the discussion by \citet{Melosh1985b}, which is that the excavation flow is driven by the residual velocity following a shock-release cycle. 
The peak-particle velocity distribution in geological media has been investigated by means of large-scale nuclear explosions. 
In this study, we used the data compilation of \citet{Perret1975}. 
The peak-particle velocity distribution, $u_{\rm p, max}(r)$, as a function of distance from the impact point $r$, is as follows \citep[e.g.,][]{Croft1982}:
\begin{eqnarray}
	u_{\rm p, max}(r)&=& u_{{\rm p}0} \quad (r<R_{\rm p}); \label{eq:u_pmax_1}\\
	u_{\rm p, max}(r)&=& u_{{\rm p}0}\left(\frac{r}{R_{\rm p}}\right)^{-n} \quad (r>R_{\rm p}), \label{eq:u_pmax_2}
\end{eqnarray}
where $u_{\rm p,max}$, $r$, $u_{{\rm p}0}$, $R_{\rm p}$, and $n$ are the peak particle velocity at a given position, the distance from the impact point, the peak-particle velocity in an isobaric core, the projectile radius, and the shock decay exponent, respectively. 
For alluvium, tuff, and granite, a shock decay exponent of $n = 1.87$ has been reported for $u_{\rm p,max}$ between $0.03$ and $5\ {\rm km\ s^{-1}}$ \citep{Perret1975, Melosh1984, Melosh1989}. 
Since $n = 1.87$ is valid for both igneous and sedimentary rocks, \citet{Melosh1984} used the value as a universal value. 
Hence, we employed this value as throughout this manuscript by following \citet{Melosh1984}.
The one-dimensional impedance matching solution has been widely used to estimate $u_{{\rm p}0}$ under a given impact condition \citep[e.g.,][]{Melosh1989}. 
If we consider a collision between two identical bodies, $u_{{\rm p}0}$ becomes half of $v_{\rm imp}$. 
Although $u_{{\rm p}0}$ depends on shock Hugoniot parameters, including the reference density $\rho_0$, the bulk sound speed $C_0$, and a constant $s$ pertaining to collisions between two different materials, $u_{{\rm p}0}$ is linearly proportional to $v_{\rm imp}$. 
The peak-particle-velocity distribution, Eq.\ (\ref{eq:u_pmax_2}), is rather similar to the radial component of the particle velocity in the $Z$-model; see Eq. (\ref{eq:u_r}). 
This peak-particle-velocity distribution, however, does not predict any excavation flow, because the exponent $n$ is less than $2$.

After the shock wave's passage, a rarefaction wave propagates into the compressed materials from the free surface. 
The compressed materials expand toward the free surface following the arrival of the rarefaction wave. 
This expansion is broadly approximated as an adiabatic process; i.e., $dS = 0$, where $S$ is the entropy. 
The adiabatic expansion is physically the same as the propagation of an expansion wave from the free surface. 
The propagating direction of the expansion wave is similar to that of the shock wave far from the impact point \citep[e.g.,][]{Melosh1985b, Kurosawa2018}. 
Since the compressed materials are accelerated into the opposite direction of the propagating expansion wave, the pressure release causes a deceleration of the compressed materials from $u_{\rm p,max}$ to $u_{\rm p,res}$, where $u_{\rm p,res}$ is the residual velocity after the pressure release. 
The magnitude of change in the particle velocity during adiabatic expansion $\Delta u_{\rm p}$ can be calculated by integrating the Riemann invariant along the isentrope \citep{Melosh1985b}:
\begin{eqnarray}
	&&\Delta u_{\rm p} = \int_{\rho_{\rm H}}^{\rho_0} \frac{C_{\rm R}}{\rho} d\rho \label{eq:Delta_u_p}\\
	&&C_{\rm R} = \sqrt{\left. \frac{\partial P}{\partial \rho}\right|_S} \label{eq:C_R}
\end{eqnarray}
where $\rho_0$, $\rho_{\rm H}$, $\rho$, $P$, and $C_{\rm R}$ are the reference density, the density in the peak shock state, the density during expansion, the pressure during expansion, and the sound speed in the compressed matter, respectively. 
The decreases in pressure and density during isentropic release are constrained by the first and second laws of thermodynamics; i.e.,
\begin{equation}
	dE = \frac{P(E,\rho)}{\rho^2}d\rho, \label{eq:dE_drho}
\end{equation}
where $E$ is the internal energy during expansion. 
The pressure $P(E, \rho)$  in Eq. (\ref{eq:dE_drho}) is given by the EOS. 
In this study, the Tillotson EOS \citep{Tillotson1962} was used to integrate Eq.\ (\ref{eq:Delta_u_p}) by combination with Eqs.\ (\ref{eq:C_R}) and (\ref{eq:dE_drho}). 
The pressure, as a function of both internal energy $E$ and density $\rho$ is given by 
\begin{equation}
	P = P_{\rm thermal}(E,\rho) + P_{\rm cold}(\rho),
\end{equation}
where $P_{\rm thermal}$ and $P_{\rm cold}$ are the thermal and cold pressures, respectively. 
The explicit expressions of $P_{\rm thermal}$ and $P_{\rm cold}$ are included in \ref{sec:A1}. 
The velocity change $\Delta u_{\rm p}$ is slightly smaller than $u_{\rm p, max}$ because of the entropy increase due to the irreversible shock heating \citep{Melosh1985b}. 
If we assume that the propagating directions of both the shock and expansion waves are the same, which is a reasonable assumption far from the impact point, the magnitude of the residual velocity after a shock-release cycle $u_{\rm p,res}$ is given by
\begin{equation}
	u_{\rm p, res}=u_{\rm p, max}-\Delta u_{\rm p}.\label{eq:upres_upmax_deltaup}
\end{equation}

\begin{figure}[htbp]
	\begin{center}
		\includegraphics[width=\linewidth]{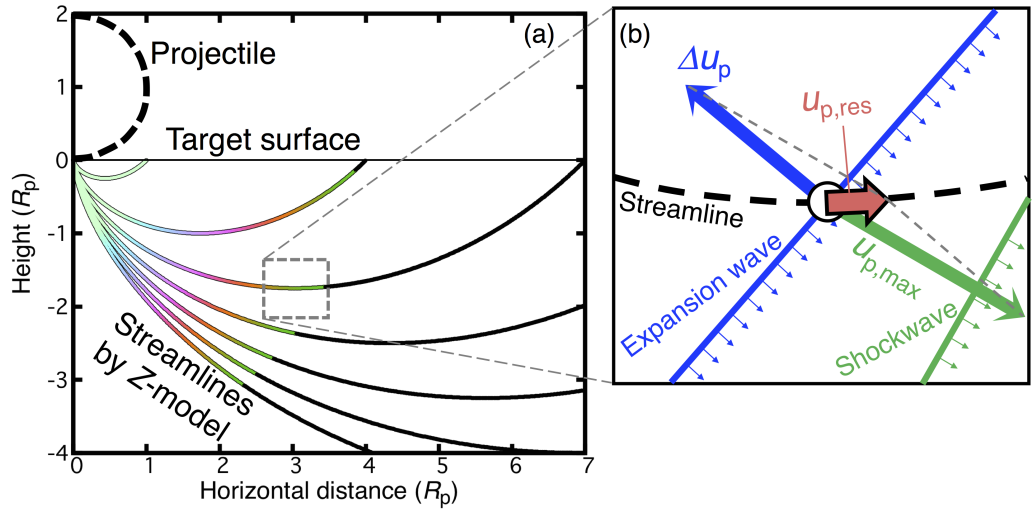}
	\end{center}
	\caption{Schematic diagram of the situation considered in the proposed model. 
	(a) The streamlines are schematically highlighted in color depending of the residual velocity. 
	Black color indicates that the residual velocity is zero; i.e., subsonic conditions, where the compressional wave speed $U$ is slower than the longitudinal sound speed $C_{\rm L}$ (See Section \ref{sec:3.1}). 
	(b) Close-up of the area indicated by the rectangle in (a). 
	The directions and magnitudes of the velocity vectors $u_{\rm p,max}$, $\Delta u_{\rm p}$, and $u_{\rm p,res}$ are shown schematically in panel (b). 
	The directions of the propagating shock and expansion waves are also shown.}
	\label{fig:fig3}
\end{figure}

The direction of $u_{\rm p,res}$ is mostly different from that of $u_{\rm p,max}$ because of the subtly different propagating directions of the shock and expansion waves \citep[e.g.,][]{Melosh1989}. 
Here, we introduce a key assumption to solve the system of equations, which is that the materials following a shock-release cycle are injected into an excavation flow along a streamline at velocity $u_{\rm p,res}$, as shown in Figure \ref{fig:fig3}. 
Figure \ref{fig:fig3}a is a schematic diagram of the cratering flow field assumed in our model. 
We assumed that the shock-driven material movement during a shock-release cycle, which occurs typically within $1\ t_{\rm s}$ (where $t_{\rm s} = D_{\rm p}/v_{\rm imp}$ is the characteristic timescale for projectile penetration), can be neglected. In other words, the shocked materials attain a particle velocity $u_{\rm p,res}$ after pressure release from their initial positions. These assumptions allow us to analytically estimate the kinetic energy available to drive material ejection in a given streamtube, as discussed in the next section. The spatial distributions of the pressure and density in the initial-peak-shock state after a single impact event, which provide the initial conditions for the integration of Eq.\ (\ref{eq:Delta_u_p}), are calculated based on the Rankine-Hugoniot relations and Eqs.\ (\ref{eq:u_pmax_1}) and (\ref{eq:u_pmax_2}) for a given impact condition.

Here, we note about the difference between impact spallation and normal excavation. 
The resultant particle velocity after a shock-release cycle strongly depends on the geometric configuration, especially on the angle between the propagation directions of the shock and expansion waves \citep{Kurosawa2018}. 
For materials initially located near the free surface, Eq.\ (\ref{eq:upres_upmax_deltaup}) does not hold, because the angle between two waves becomes $\sim 90^\circ$, resulting in high-speed ejecta caused by impact spallation \citep{Kurosawa2018}. 
Since the mass ejected by spallation is estimated to be much smaller than that from normal excavation considered in this study \citep[e.g.,][]{Melosh1984, Kurosawa2018}, we neglected the near-surface wave interactions.

\subsection{Impact ejection and cessation of crater growth}\label{sec:2.3}
Given that the shapes of the streamlines and the kinetic energy in a streamtube have been obtained, we can now assess the energy balance of the kinetic, gravitational, and strength energies ($E_{\rm kin}$, $E_{\rm grav}$, and $E_{\rm strength}$, respectively) at a given horizontal distance $R$. 
The strength energy is the energy required to move materials supported by a yield strength $Y$. 
The mass of a streamtube $M_{\rm tube}$ between $R-{\it \Delta} R$ and $R$ is calculated using the following volume integral in polar coordinates:
\begin{equation}
	M_{\rm tube}(R) = 2\pi \rho_{\rm t} \int_{R-{\it \Delta} R}^R \int_0^{\frac{\pi}{2}} r^2 \sin\theta drd\theta,\label{eq:M_tube}
\end{equation}
where ${\it \Delta} R$ and $\rho_{\rm t}$ are a small increment in the horizontal distance and the reference density of the target, respectively. 
$E_{\rm kin}$, $E_{\rm grav}$, and $E_{\rm strength}$ are expressed as
\begin{eqnarray}
	E_{\rm kin}(R) &=& 2\pi \rho_{\rm t} \int_{R-{\it \Delta} R}^R \int_0^{\frac{\pi}{2}} \frac{u_{\rm p, res}^2}{2} r^2\sin\theta dr d\theta,\\
	E_{\rm grav}(R) &=& 2\pi \rho_{\rm t} \int_{R-{\it \Delta} R}^R \int_0^{\frac{\pi}{2}} gz r^2\sin\theta dr d\theta,
\end{eqnarray}
and
\begin{equation}
	E_{\rm strength} (R)= 2\pi \int_{R-{\it \Delta} R}^R \int_0^{\frac{\pi}{2}} \varepsilon Y r^2\sin\theta dr d\theta,\label{eq:M_strength}
\end{equation}
where $z = –r\cos\theta$ is the height from the pre-impact surface and $\varepsilon$ is the volumetric strain. 
If $E_{\rm kin}$ is greater than the sum of $E_{\rm grav}$ and $E_{\rm strength}$ at a given distance $R$, the materials in the streamtube are ejected. 
The ejection velocity $v_{\rm ej}$ is estimated from energy conservation as
\begin{equation}
	v_{\rm ej}= \sqrt{\frac{2(E_{\rm kin} - E_{\rm grav} - E_{\rm strength})}{M_{\rm tube}}}. \label{eq:v_ej}
\end{equation}

Given that the characteristic velocity of the material in each streamtube $v_{\rm ch}$ is roughly approximated by $v_{\rm ch}\sim v_{\rm ej}$, the time variation in the radius of a growing crater $R_{\rm cr}(t)$ can be obtained as a first-order estimate. 
The time $t_{\rm ej}$ when ejection occurs, from position $R$, is estimated as 
\begin{equation}
	t_{\rm ej}= \frac{L}{v_{\rm ch}},
\end{equation}
where $L$ is the total travel distance along the streamline, and $L = f(Z)R$, where $f(Z)$ is a constant that depends only on $Z$. 
The exact form of $f(Z)$ is presented in \ref{sec:A2}. 
The time variation of $R_{\rm cr}(t)$ is given by
\begin{equation}
	R_{\rm cr}(t) = \frac{v_{\rm ch}}{f(Z)}t_{\rm ej}. \label{eq:R_cr}
\end{equation}
Cessation of the growth of a crater occurs when $E_{\rm kin} = E_{\rm grav}$ or $E_{\rm kin} = E_{\rm strength}$ in the gravity- or strength-dominated regimes, respectively. 
These conditions provide an absolute value of the transient crater radius $R_{\rm tr}$ under a given impact condition.

Most of the equations described in this section can be solved analytically if the shock decay exponent $n$, the Tillotson and shock Hugoniot parameters, and the basic quantities $v_{\rm imp}$, $D_{\rm p}$, $g$, $Y$, $\rho_{\rm p}$, and $\rho_{\rm t}$ are known. 
The exceptions can also be integrated easily using a spreadsheet. 
Consequently, our model can be used to estimate the absolute value of $R_{\rm tr}$ analytically without reliance on the $\pi$-group scaling laws.

It should be mentioned that \citet{OKeefe1981}, \citet{Ivanov1983}, and \citet{Richardson2007} proposed similar analytical models, although with key differences to the present model. 
\citet{OKeefe1981} addressed the controls on transient crater depth, not the crater diameter, and did not include the effects of the residual velocity in their model. 
\citet{Ivanov1983} also employed the geometry predicted by the $Z$-model to calculate mechanical plastic work in ideal plastic media with a constant yield strength. 
He estimated the decay of the ejection velocity with respect to the horizontal distance from the impact point in the strength-dominated regime. 
The model proposed by \citet{Richardson2007} was constructed by means of a combination of Maxwell's $Z$-model and point-source theory to systematically investigate the effects of gravitational acceleration, target density, and target strength on ejection behavior in the framework of the $\pi$-group scaling laws.

\section{Results}
In this section, we present explicit expressions for the variables described in Section \ref{sec:2} as well as the results of our calculations. 
First, we discuss the residual velocity $u_{\rm p,res}$ as a function of the peak particle velocity behind the shock wave $u_{\rm p,max}$ (Section \ref{sec:3.1}). 
Second, in Section \ref{sec:3.2} we show the integrated results of the energies in a streamtube. 
Third, the ejection behavior, including the time evolution of the radius of a growing crater and the ejection velocity distribution, is described in Section \ref{sec:3.3}. 
Finally, we present the resultant crater sizes $R_{\rm tr}$ in Section \ref{sec:3.4}.

\subsection{Residual velocity}\label{sec:3.1}
\begin{figure}[htbp]
	\begin{center}
		\includegraphics[width=\linewidth]{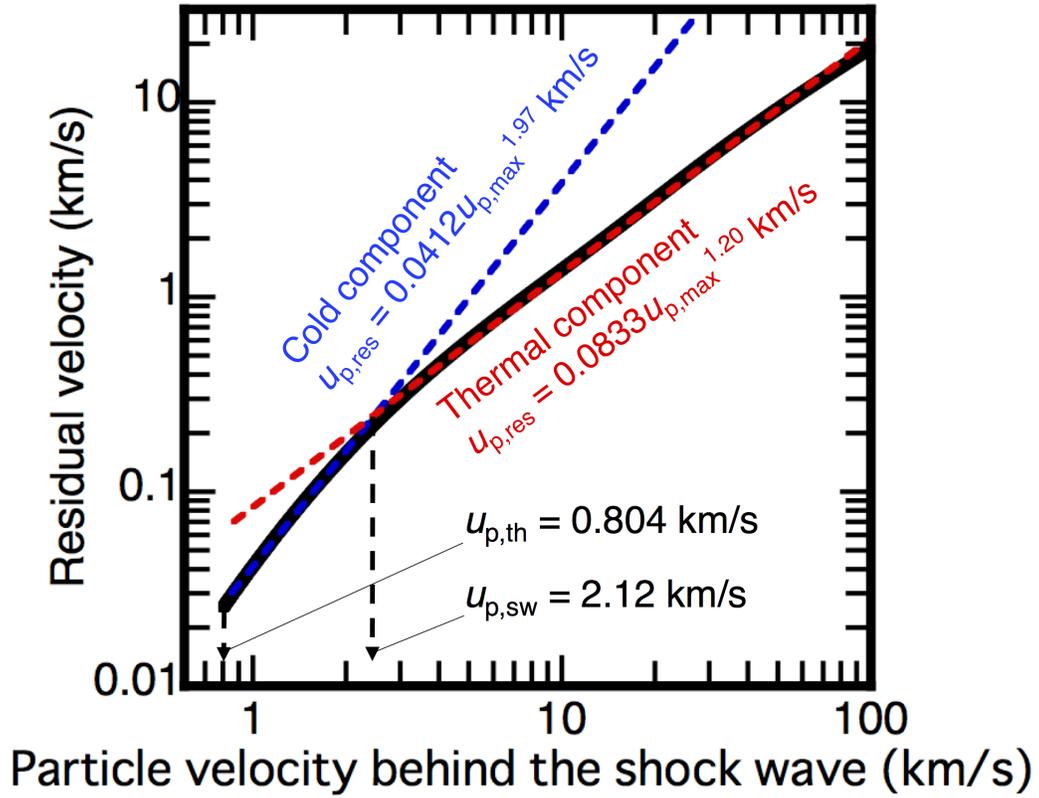}
	\end{center}
	\caption{Residual velocity $u_{\rm p,res}$ as a function of the peak particle velocity $u_{\rm p,max}$ pertaining to granite. 
	Two power-law functions, referred to as the thermal and cold components, are shown as dotted lines. 
	The values of the switching and threshold velocities (see Section \ref{sec:3.1}) are also included.}
	\label{fig:fig4}
\end{figure}

\begin{figure}[htbp]
	\begin{center}
		\includegraphics[width=\linewidth]{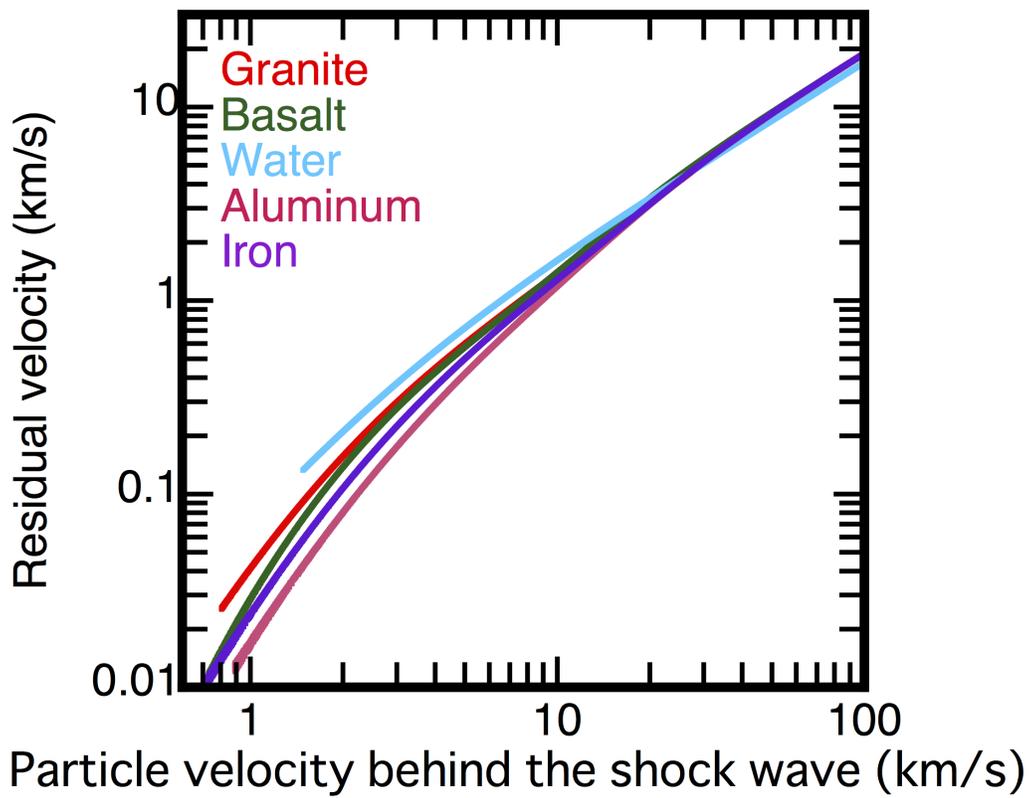}
	\end{center}
	\caption{As for Fig. \ref{fig:fig4}, but showing the relation between the residual velocity $u_{\rm p,res}$ and the peak particle velocity $u_{\rm p,max}$ for various materials.}
	\label{fig:fig5}
\end{figure}

We calculated the residual velocity $u_{\rm p,res}$ using Eqs.\ (\ref{eq:u_pmax_1})--(\ref{eq:upres_upmax_deltaup}). 
Figure \ref{fig:fig4} shows $u_{\rm p,res}$ as a function of $u_{\rm p,max}$ for granite. 
We found that $u_{\rm p,res}$ is approximated by two power-law functions with coefficients $C$ ($C_{\rm c}$ and $C_{\rm t}$) and exponents $m$ ($m_{\rm c}$ and $m_{\rm t}$), as follows: 
\begin{equation}
	u_{\rm p, res}(r) = C u_{\rm p, max}(r)^m, \label{eq:u_pres}
\end{equation}
where
\begin{eqnarray}
	&&C=C_{\rm t}\ {\rm and}\ m=m_{\rm t} \quad  (u_{\rm p, max}(r)> u_{\rm p, sw}),\\
	&&C=C_{\rm c}\ {\rm and}\ m=m_{\rm c} \quad (u_{\rm p, th}<u_{\rm p, max}(r)< u_{\rm p, sw}),
\end{eqnarray}
where $u_{\rm p, sw}$ and $u_{\rm p, th}$ are switching and threshold velocities, respectively. 
Note that the subscripts ``t'' and ``c'' mean `thermal' and `cold,' respectively. 
The switching velocity corresponds to the transition from the cold-pressure-dominated regime ($P_{\rm cold} > P_{\rm thermal}$) to its thermal-pressure-dominated counterpart ($P_{\rm cold} < P_{\rm thermal}$) with increasing $u_{\rm p,max}$. 
The threshold velocity corresponds to the transition from the elastic-plastic state to the shocked state in a diagram showing the wave speed $U$ versus the particle velocity $u_{\rm p}$. 
The definition of $u_{\rm p, th}$ is provided in \ref{sec:A3}. 
The absolute magnitude of the residual velocity is estimated to range from $4\%$ to $20\%$ of that of the peak particle velocity. 
A higher $u_{\rm p,max}$ leads to a higher $u_{\rm p,res}$ because the shock-induced entropy is enhanced at higher shock pressures. 
Figure \ref{fig:fig5} is the same as Figure \ref{fig:fig4}, except that the former shows the $u_{\rm p,res} - u_{\rm p,max}$ relation pertaining to various materials. 
The Tillotson parameters used in the calculations were taken from \citet{Melosh1989} and \citet{Benz1999}. 
These results suggest that the combination of two power laws leads to a unified equation to describe the behavior of $u_{\rm p,res}$ with increasing $u_{\rm p,max}$. 
The best-fitting values for $C$ and $m$ are listed in Table \ref{fig:table}. 
We checked the absolute magnitude of the residual velocity based on a comparison with a hydrocode. 
The limitations of our model are described in Section \ref{sec:4.1}. 

\begin{table}[htb]
	\begin{center}
		\caption{Parameters relating to the residual velocity}
		\label{fig:Table1}
 		\begin{tabular}{|l||c|c|c|c|c|} \hline
   			 & Granite & Basalt & Water & Aluminum & Iron \\ \hline \hline
    			$u_{\rm p, th}^1$ (km/s) & 0.804 & 0.730 & 1.48${}^2$ & 0.887 & 0.652 \\ \hline
    			$u_{\rm p, sw}$ (km/s) & 2.12 & 2.67 & 5.03 & 5.45 & 3.19 \\ \hline
    			$C_{\rm c}$ & 0.0412 & 0.0281 & 0.0796 & 0.0187 & 0.0235 \\ \hline
    			$m_{\rm c}$ & 1.97 & 2.29 & 1.39 & 2.00 & 2.14 \\ \hline
    			$nm_{\rm c}^3$ & 3.69 & 4.28 & 2.60 & 3.73 & 4.00 \\ \hline
			$C_{\rm t}$ & 0.0833 & 0.0827 & 0.145 & 0.0655 & 0.0691 \\ \hline
			$m_{\rm t}$ & 1.20 & 1.21 & 1.04 & 1.26 & 1.25 \\ \hline
			$nm_{\rm t}$ & 2.25 & 2.27 & 1.95 & 2.36 & 2.34 \\ \hline
  		\end{tabular}
  	\end{center}
  	${}^1$The parameters $C_0$ and $s$ are taken from \citet{Melosh1989}. 
  	The longitudinal sound speeds $C_{\rm L}$ are calculated using the Poisson ratio and $C_0$.\\
  	${}^2$Since liquid water is considered here, $C_{\rm L}$ corresponds to the bulk sound speed $C_0$.\\
	${}^3$The decay exponent $n$ is assumed to be constant, $n = 1.87$ (see Section \ref{sec:2.2}).
	\label{fig:table}
\end{table}

The peak particle velocity $u_{\rm p,max}$ is also expressed as a power-law function [see Eq.\ (\ref{eq:u_pmax_2})] with respect to the distance from the impact point $r$ (for $r > R_{\rm p}$). 
Thus, $u_{\rm p,res}$ can be rewritten from Eq.\ (\ref{eq:u_pres}) as a function of $r$, as follows:
\begin{equation}
	u_{\rm p, res} = C u_{{\rm p}0}^m \left(\frac{r}{R_{\rm p}}\right)^{-nm} \quad (r> R_{\rm p}). \label{eq;u_pres_2}
\end{equation}
Analogously to the Maxwell $Z$-model, the exponent $Z$ is related to the product of the exponents $n$ and $m$. 
Since $n = 1.87$ (see Section \ref{sec:2.2}) and $m > 1.2$ in most cases, except for $m_{\rm t}$ for water (Table \ref{fig:table}), our model naturally reproduces $Z > 2$, which is the key constraint to describe streamlines (Section \ref{sec:2.1}). 
Although we independently employed the values $nm$ and $Z$ from thermodynamic considerations [Eq.\ (\ref{eq;u_pres_2})] and Maxwell $Z$-model [(Eq. (\ref{eq:u_r})), respectively, they are not mutually exclusive.

\subsection{Energies in stream tubes}\label{sec:3.2}
\begin{figure}[htbp]
	\begin{center}
		\includegraphics[width=\linewidth]{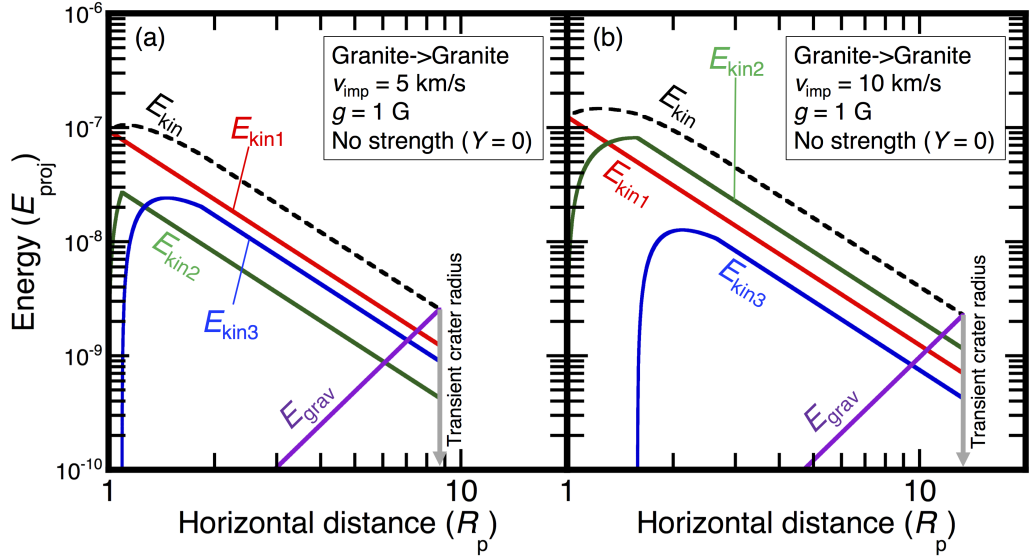}
	\end{center}
	\caption{Horizontal profiles of kinetic energy and gravitational potential energy in a given streamtube. 
	The impact velocity is (a) $5\ {\rm km\ s^{-1}}$ and (b) $10\ {\rm km\ s^{-1}}$. 
	The sum of the kinetic energies $E_{\rm kin}$ (black dashed line) of the thermal components $E_{{\rm kin}1}$ (red solid line) and $E_{{\rm kin}2}$ (green solid line), the cold component $E_{{\rm kin}3}$ (blue solid line), as well as the gravitational potential energy $E_{\rm grav}$ (purple solid line) are plotted. 
	The horizontal distance from the impact point $R$ and the energies is normalized by the projectile radius $R_{\rm p}$ and the kinetic energy of the projectile $E_{\rm proj}$.}
	\label{fig:fig6}
\end{figure}

In this section, we describe the integrated results of Eqs.\ (\ref{eq:M_tube})--(\ref{eq:M_strength}). 
Note that here we only show the proportional relations with dimensional constants from $k_1$ to $k_9$. 
The expressions of the coefficients are presented in \ref{sec:A4}. 
The total mass in a given streamtube is as follows \citep[e.g.,][]{Maxwell1977}:
\begin{equation}
	M_{\rm tube}=k_1 R^2 {\it \Delta} R,
\end{equation}
When $u_{{\rm p}0} > u_{\rm p, sw}$, $E_{\rm kin}$ is divided into three terms:
\begin{equation}
	E_{\rm kin}=E_{{\rm kin}1}+E_{{\rm kin}2}+E_{{\rm kin}3}, \label{eq:E_kin}
\end{equation}
where
\begin{eqnarray}
	E_{{\rm kin}1} &=& k_2 u_{{\rm p}0}^{2m_{\rm t}} R_{\rm p}^{Z+1} R^{-(Z-1)}{\it \Delta} R, \label{eq:E_kin1}\\
	E_{{\rm kin}2} &=& \left(k_3 u_{{\rm p}0}^{2m_{\rm t}} + k_4 u_{{\rm p}0}^{\frac{Z+1}{n}}\right) R_{\rm p}^{Z+1} R^{-(Z-1)}{\it \Delta}R,
\end{eqnarray}
and
\begin{equation}
	E_{{\rm kin}3} = k_5 u_{{\rm p}0}^{\frac{Z+1}{n}} R_{\rm p}^{Z+1} R^{-(Z-1)}{\it \Delta} R.\label{eq:E_kin3}
\end{equation}
The contribution of the kinetic energy inside the isobaric core to the streamtube corresponds to $E_{{\rm kin}1}$. 
The second and third terms ($E_{{\rm kin}2}$ and $E_{{\rm kin}3}$) originate mainly from the thermal pressure $P_{\rm thermal}$ (thermal component) and the cold pressure $P_{\rm cold}$ (cold component), respectively, on the outside of the isobaric core. 
Since $u_{{\rm p}0} > u_{\rm p,sw}$ was assumed in Eqs.\ (\ref{eq:E_kin1})--(\ref{eq:E_kin3}), the kinetic energy of the material initially located inside the isobaric core $E_{{\rm kin}1}$ is also classified as a thermal component. 
In contrast, if $u_{{\rm p}0}$ is slower than $u_{\rm p,sw}$, the thermal component disappears:
\begin{equation}
	E_{\rm kin}=E_{{\rm kin}1}+E_{{\rm kin}2},\label{eq:E_kin_}
\end{equation}
where
\begin{eqnarray}
	E_{{\rm kin}1} &=& k_6 u_{{\rm p}0}^{2m_{\rm c}} R_{\rm p}^{Z+1} R^{-(Z-1)}{\it \Delta} R, \\
	E_{{\rm kin}2} &=& \left(k_7 u_{{\rm p}0}^{2m_{\rm c}} + k_8 u_{{\rm p}0}^{\frac{Z+1}{n}}\right) R_{\rm p}^{Z+1} R^{-(Z-1)}{\it \Delta} R.\label{eq:E_kin2_}
\end{eqnarray}
Since $u_{{\rm p}0}$ is linearly proportional to $v_{\rm imp}$, as discussed in Section \ref{sec:2.2}, Eqs.\ (\ref{eq:E_kin})--(\ref{eq:E_kin3}) and (\ref{eq:E_kin_})--(\ref{eq:E_kin2_}) also include the dependence of $v_{\rm imp}$ on $E_{\rm kin}$. 
The gravitational potential energy of a given streamtube $E_{\rm grav}$ is expressed as
\begin{equation}
	E_{\rm grav}=k_9 gR^3{\it \Delta} R. \label{eq:E_grav}
\end{equation}
Figure \ref{fig:fig6} displays typical examples of the calculations, showing the kinetic energy and the gravitational potential energy in a given streamtube as a function of the horizontal distance $R$. 
A granite projectile and target were considered in these calculations. 
We set $v_{\rm imp}$ at $5\ {\rm km\ s^{-1}}$ and $10\ {\rm km\ s^{-1}}$ in panels (a) and (b), respectively. 
To draw the figures, $Z$, $g$, and ${\it \Delta} R$ were set at $3$, $1\ G$, and $10^{-4}\ R_{\rm p}$, respectively. 
Since the $u_{{\rm p}0}$ velocities are higher than $u_{\rm p,sw}$ in both cases, $E_{{\rm kin}1}$ and $E_{{\rm kin}2}$ are associated with the thermal component under the calculation conditions. 
All kinetic energy terms decrease following the power law, $E_{\rm kin}\propto R^{-2}$: see Eqs.\ (\ref{eq:E_kin1})--(\ref{eq:E_kin3}). 
$E_{\rm kin}$ and $E_{\rm grav}$ are balanced at $8.7\ R_{\rm p}$ and $13.2\ R_{\rm p}$ for $v_{\rm imp} = 5\ {\rm km\ s^{-1}}$ and $10\ {\rm km\ s^{-1}}$, respectively. 
Although the gravitational potential energy increases more steeply with increasing $R$ ($E_{\rm grav}\propto R^3$; Eq.\ (\ref{eq:E_grav})), the deceleration owing to gravity can be neglected until the radius of the growing crater, $R_{\rm cr}(t)$, approaches the transient crater radius. 
As discussed in the next section, $v_{\rm ej}(R)$ and $R_{\rm cr}(t)$ in the region where $E_{\rm grav}$ can be neglected exhibit the well-known power-law behavior. 

Given that the yield strength is a constant and that the volumetric strain is approximated by unity as a first-order estimate, $E_{\rm strength}$ is roughly approximated by
\begin{equation}
	E_{\rm strength} = \frac{M_{\rm tube}}{\rho_{\rm t}} Y = k_1\left(\frac{Y}{\rho_{\rm t}}\right) R^2 {\it \Delta} R.
\end{equation}
Strictly speaking, the treatment of the material strength used above is affected by three main problems, as described in the remainder of this section. 
The first problem is that we neglect the effects of the velocity difference between the adjacent stream tubes to derive Eq.\ (\ref{eq:v_ej}). 
The volumetric strain $\varepsilon$ is essentially computed from the velocity difference. 
Thus, we assumed that the volumetric strain $\varepsilon=1$ to calculate $E_{\rm strength}$ is a first-order estimate, as mentioned above. 
The second problem is related to the first one; i.e., our neglect of the frictional behavior of geologic materials. 
Actual geologic media exhibit a complicated strength behavior because of dry friction \citep[e.g.,][]{Lundborg1968}. 
If the work done by the frictional drag force greatly affects the energy partitioning of the excavation flow, we cannot apply Eq.\ (\ref{eq:v_ej}) to estimate the ejection velocity. 
Although we could in principle estimate the significance of the friction using numerical simulations with constitutive models, such advanced calculations are beyond the scope of this study. 
The final problem is the constant-$Y$ assumption, which is only valid for metal-like targets. 
Despite this limitation, the constant-$Y$ assumption has been widely used to derive the $\pi$-group scaling laws \citep[e.g.,][]{Gault1973, Suzuki2012}. 
Thus, we decided to present the cratering processes in the strength-dominated regime based on the constant-$Y$ assumption. 
Because of these three simplifications, the predictions of our model regarding the impact outcomes in the strength-dominated regime are expected to be associated with significant uncertainties in this regime.


\begin{figure}[htbp]
	\begin{center}
		\includegraphics[width=\linewidth]{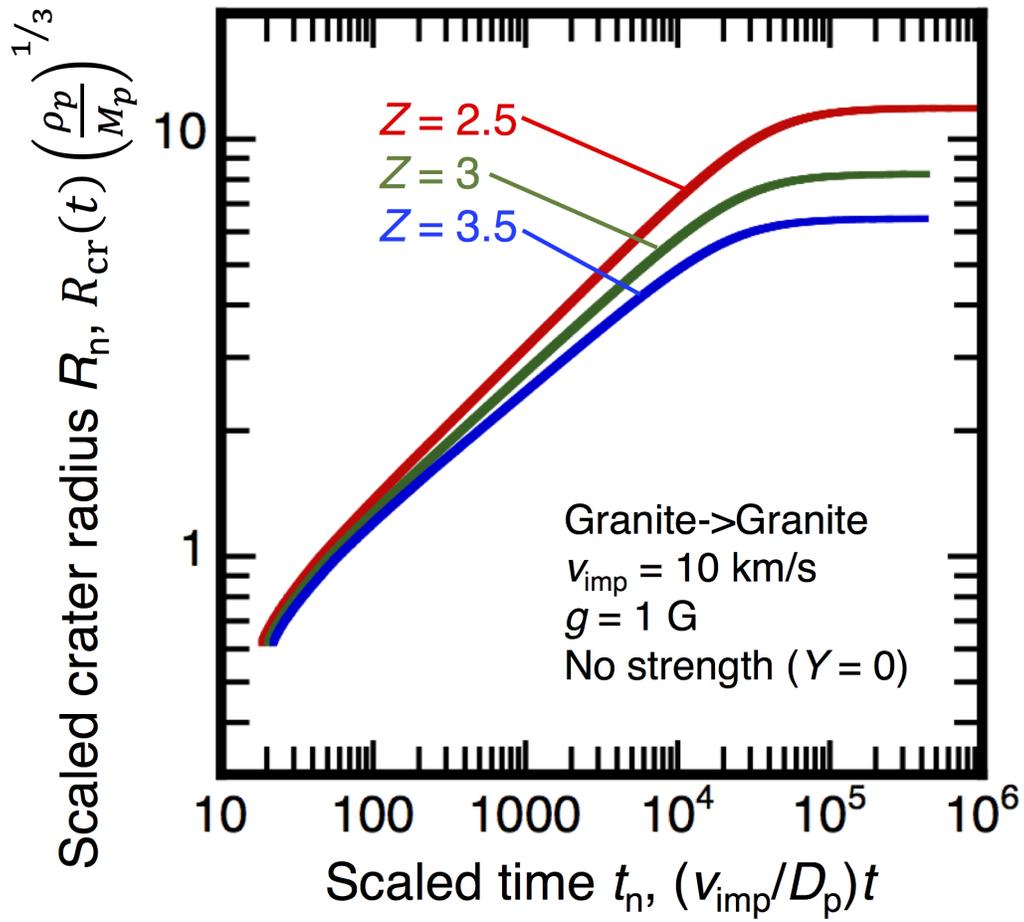}
	\end{center}
	\caption{Time variation of the radii of growing craters. 
	The relevant conditions are listed in the figure. 
	Both the crater radii and the time after impact are normalized so as to be dimensionless variables. 
	Results are shown for $Z = 2.5$ (red), $3.0$ (green), and $3.5$ (blue).}
	\label{fig:fig7}
\end{figure}

\subsection{Ejection behavior}\label{sec:3.3}
\begin{figure}[htbp]
	\begin{center}
		\includegraphics[width=\linewidth]{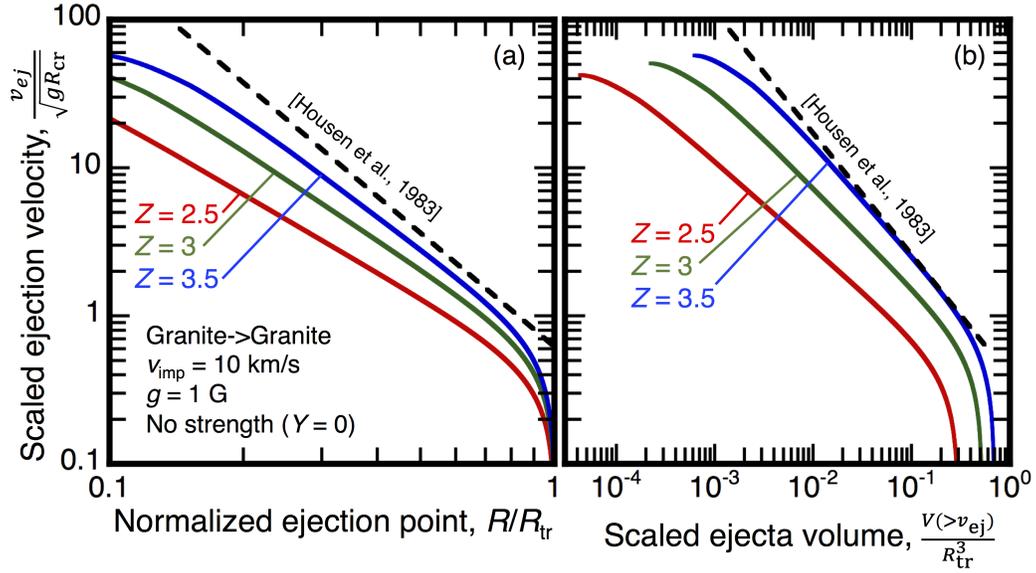}
	\end{center}
	\caption{(a) Ejecta velocity distribution and (b) relation between the cumulative volume of the ejecta launched at a given ejection velocity and the ejection velocity. 
	These results are obtained from the same calculation as that shown in Fig.\ \ref{fig:fig7}. 
	The previous ejecta scaling laws of \citet{Housen1983}, (a) $\frac{v_{\rm ej}}{\sqrt{gR_{\rm tr}}}=0.62 \left(\frac{R}{R_{\rm tr}}\right)^{-2.55}$ and (b) $\frac{V(>v_{\rm ej})}{R_{\rm tr}^3}=0.32\left(\frac{v_{\rm ej}}{\sqrt{gR_{\rm tr}}}\right)^{-1.22}$, are also shown (dotted lines).}
	\label{fig:fig8}
\end{figure}

In this section, we discuss the ejecta characteristics pertaining to our model. 
Figure \ref{fig:fig7} shows examples of crater growth in the gravity-dominated regime ($E_{\rm grav} \gg E_{\rm strength}$) calculated using Eqs.\ (\ref{eq:v_ej})--(\ref{eq:R_cr}), (\ref{eq:E_kin}), and (\ref{eq:E_grav}). 
We assumed that a granite projectile of $1\ {\rm m}$ in radius collides with a strengthless granite target at $10\ {\rm km\ s^{-1}}$ under a gravitational acceleration $g = 1\ {\rm G} = 9.81\ {\rm m\ s^{-2}}$. 
Three different $Z$ values were used ($2.5$, $3.0$, and $3.5$). 
Normalized radii of the growing craters $R_{\rm n} = (\rho_{\rm p}/M_{\rm p})^{1/3}R_{\rm cr}$ are plotted against a normalized time quantity, $t_{\rm n} = t/t_{\rm s}$. 
Crater growth under these conditions follows a power law before $t_{\rm n} \simeq 10^4$. 
Next, the growth rates gradually decrease with time, and they cease around $t_{\rm n} \simeq 10^5$. 
Although the absolute value of the timing of the material ejection ($x$ axis) is a first-order estimate, as mentioned in Section \ref{sec:2.3}, because of the uncertainty in the characteristic velocity of the material in a given streamtube, the time sequence is consistent with those derived in previous experimental studies \cite[e.g.,][]{Yamamoto2009, Yamamoto2017}. 
The power-law exponent depends on $Z$. 
Our model naturally reproduces the power-law behavior, as follows. 
The ejection velocity $v_{\rm ej}$ under the condition where $E_{\rm kin} \gg E_{\rm grav}$ (and/or $E_{\rm strength}$), which actually holds in our model (see Fig.\ \ref{fig:fig6}), is practically equal to 
\begin{equation}
	v_{\rm ej} = \sqrt{\frac{2E_{\rm kin}}{M_{\rm tube}}}. \label{eq:v_ej_2}
\end{equation}
In this case, the ejection velocity distribution (i.e., $v_{\rm ej}$ as a function of $R$) and the time variation of $R_{\rm cr}$ can be expressed as power-law functions as follows:
\begin{equation}
	v_{\rm ej}\propto R^{-\frac{Z+1}{2}}, \label{eq:v_ej_R}
\end{equation}
and
\begin{equation}
	R_{\rm cr}\propto t^{\frac{2}{Z+3}}. \label{eq:R_t_Z}
\end{equation}
Figure \ref{fig:fig8} shows the ejecta velocity distributions. 
We calculated the ejection velocity and the ejecta volume launched at a higher velocity than a given ejection velocity using the same calculations as those in Fig.\ \ref{fig:fig7}. 
Following \citet{Housen1983}, we plotted the normalized ejection velocity $v_{\rm ej}/\sqrt{gR_{\rm tr}}$ as a function of both the normalized ejecta position $R/R_{\rm tr}$ (Fig.\ \ref{fig:fig8}a) and the scaled ejecta volume $V(>v_{\rm ej})/R_{\rm tr}^3$ (Fig. \ref{fig:fig8}b). 
To compare with previous results, the best-fitting lines from \citet{Housen1983}, based on point-source theory, are also plotted. 
As described by Eq.\ (\ref{eq:v_ej_R}), the scaled ejection velocities exhibit power-law behaviors that depend on $Z$ (Fig.\ \ref{fig:fig8}a). 
For $Z = 3.5$, the slope is close to the best-fit line of \citet{Housen1983}. 
The difference in the absolute value between the line for $Z = 3.5$ and the best-fit line (blue solid and dotted lines, respectively) might originate from the fact that the best-fit line was determined using the final crater radii rather than the transient crater radii. 
The line for $Z = 3.5$ is consistent with the result of \citet{Housen1983} in terms of the ejecta volume at a given ejection velocity (Fig.\ \ref{fig:fig8}b). 
The scaled ejecta volume deviates from the power law (the dotted line) in regions characterized by relatively high and low scaled ejection velocities. 
This behavior is consistent with previous numerical results obtained by \citet{Wada2006}. 
Since the validity of Eq.\ (\ref{eq:v_ej_2}) breaks down at relatively low ejection velocities (i.e., $E_{\rm kin} \sim E_{\rm grav}$), the results deviate from the power-law behavior. 
A cut-off at high ejection velocities is discussed in detail in Section \ref{sec:4.2}. 
Consequently, our model predicts a similar power-law behavior as that proposed in previous studies \citep[e.g.,][]{Housen1983, Schmidt1987}. 
The correspondence between our model and the point-source theory is discussed in Section \ref{sec:4.3}.

\subsection{Transient crater radii}\label{sec:3.4}
The transient crater radius $R_{\rm tr}$ in the gravity-dominated regime, defined as $E_{\rm grav} \gg E_{\rm strength}$, is obtained by assuming $E_{\rm kin} = E_{\rm grav}$, so that
\begin{equation}
	R_{\rm tr}= R_{\rm p}^{\frac{Z+1}{Z+2}}(k_9 g)^{-\frac{1}{Z+2}} 
		\left[(k_2+k_3)u_{{\rm p}0}^{2m_{\rm t}} +(k_4+k_5)u_{{\rm p}0}^{\frac{Z+1}{n}}\right]^{\frac{1}{Z+2}} 
		\quad ({\rm if}\ u_{{\rm p}0}>u_{\rm p, sw}),\label{eq:R_tr_kin_grav}
\end{equation}
and
\begin{equation}
	R_{\rm tr}= R_{\rm p}^{\frac{Z+1}{Z+2}}(k_9 g)^{-\frac{1}{Z+2}} 
		\left[(k_6+k_7)u_{{\rm p}0}^{2m_{\rm c}} +k_8 u_{{\rm p}0}^{\frac{Z+1}{n}}\right]^{\frac{1}{Z+2}} 
		\quad ({\rm if}\ u_{{\rm p}0}<u_{\rm p, sw}). \label{eq:R_tr_kin_grav2}
\end{equation}
In the strength-dominated regime (i.e., $E_{\rm grav} \ll E_{\rm strength}$), $R_{\rm tr}$ is calculated by assuming $E_{\rm kin} = E_{\rm strength}$, so that
\begin{equation}
	R_{\rm tr}= R_{\rm p} \left(\frac{k_1 Y}{\rho_{\rm t}}\right)^{-\frac{1}{Z+1}}
		\left[(k_2+k_3)u_{{\rm p}0}^{2m_{\rm t}} +(k_4+k_5)u_{{\rm p}0}^{\frac{Z+1}{n}}\right]^{\frac{1}{Z+1}} 
		\quad ({\rm if}\ u_{{\rm p}0}>u_{\rm p, sw}),\label{eq:R_tr_kin_strength}
\end{equation}
and
\begin{equation}
	R_{\rm tr}= R_{\rm p} \left(\frac{k_1 Y}{\rho_{\rm t}}\right)^{-\frac{1}{Z+1}}
		\left[(k_6+k_7)u_{{\rm p}0}^{2m_{\rm c}} +k_8 u_{{\rm p}0}^{\frac{Z+1}{n}}\right]^{\frac{1}{Z+1}} 
		\quad ({\rm if}\ u_{{\rm p}0}>u_{\rm p, sw}).
\end{equation}
Of note, $R_{\rm tr}$ in our model depends on the exponents $n$ and $m$, indicating that the nature of the decaying shock propagation and the thermodynamic/hydrodynamic response of geological materials are included to predict the resulting crater sizes.

\begin{figure}[htbp]
	\begin{center}
		\includegraphics[width=\linewidth]{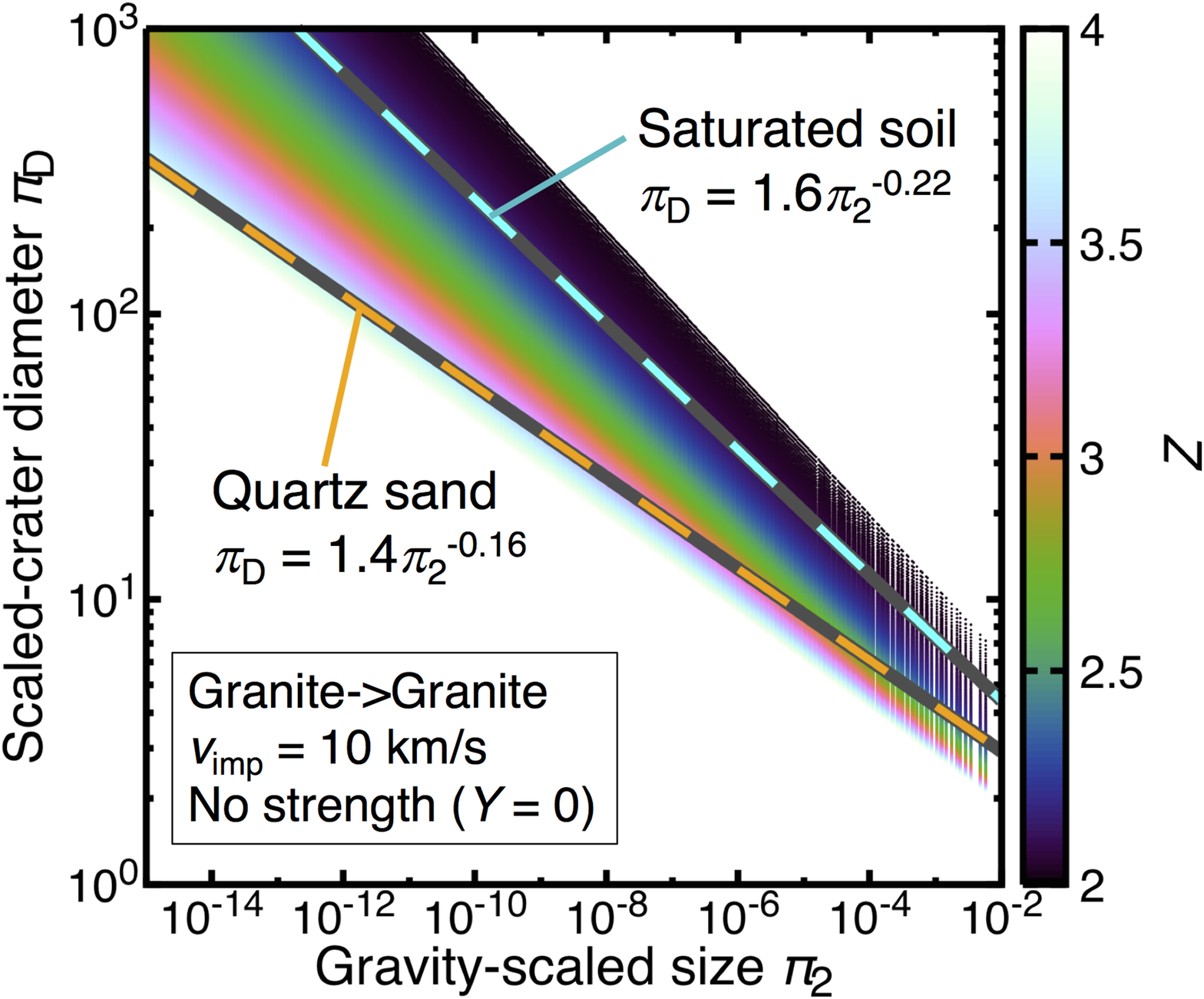}
	\end{center}
	\caption{Scaled crater diameter $\pi_D$ as a function of $\pi_2$. 
	The calculated $\pi_D$ values are colored depending on the exponent $Z$. 
	The other calculation conditions are also indicated in the figure. 
	Two typical scaling lines for saturated soil (cyan) and dry quartz sand (orange) are shown. 
	The scaling constant and exponent are from \citet{Schmidt1987}.}
	\label{fig:fig9}
\end{figure}
\begin{figure}[htbp]
	\begin{center}
		\includegraphics[width=\linewidth]{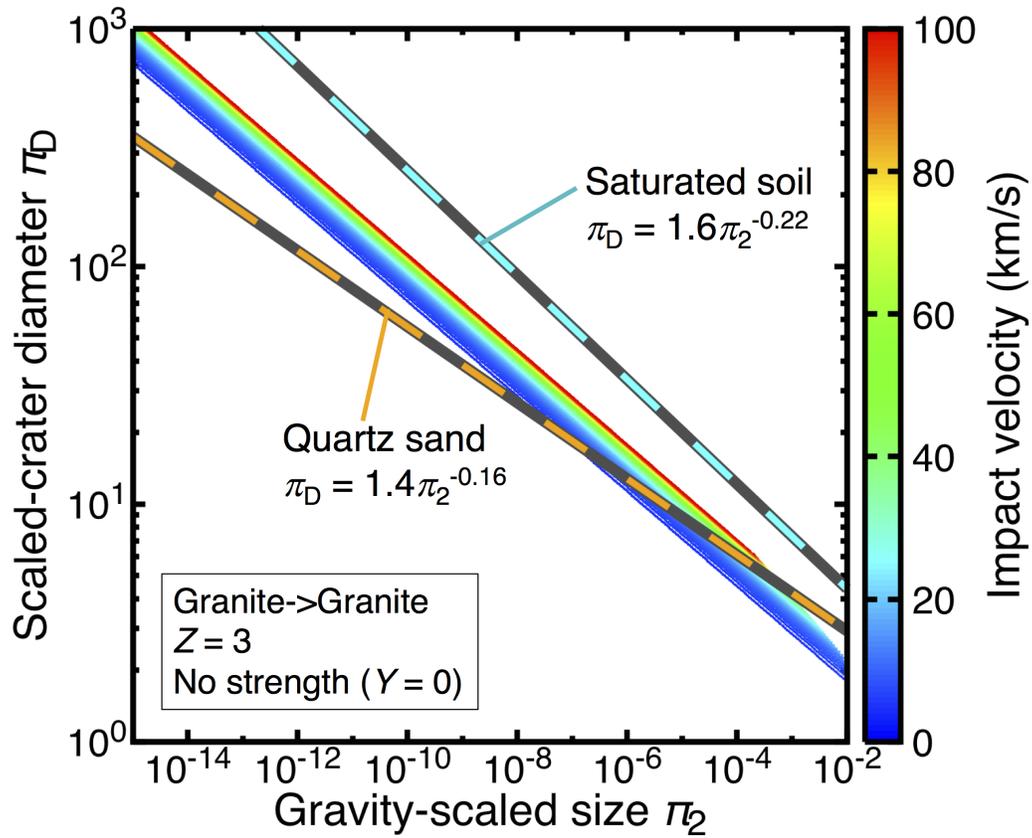}
	\end{center}
	\caption{As for Fig.\ \ref{fig:fig9}, except that the calculated $\pi_D$ values are colored depending on the impact velocity.}
	\label{fig:fig10}
\end{figure}
\begin{figure}[htbp]
	\begin{center}
		\includegraphics[width=\linewidth]{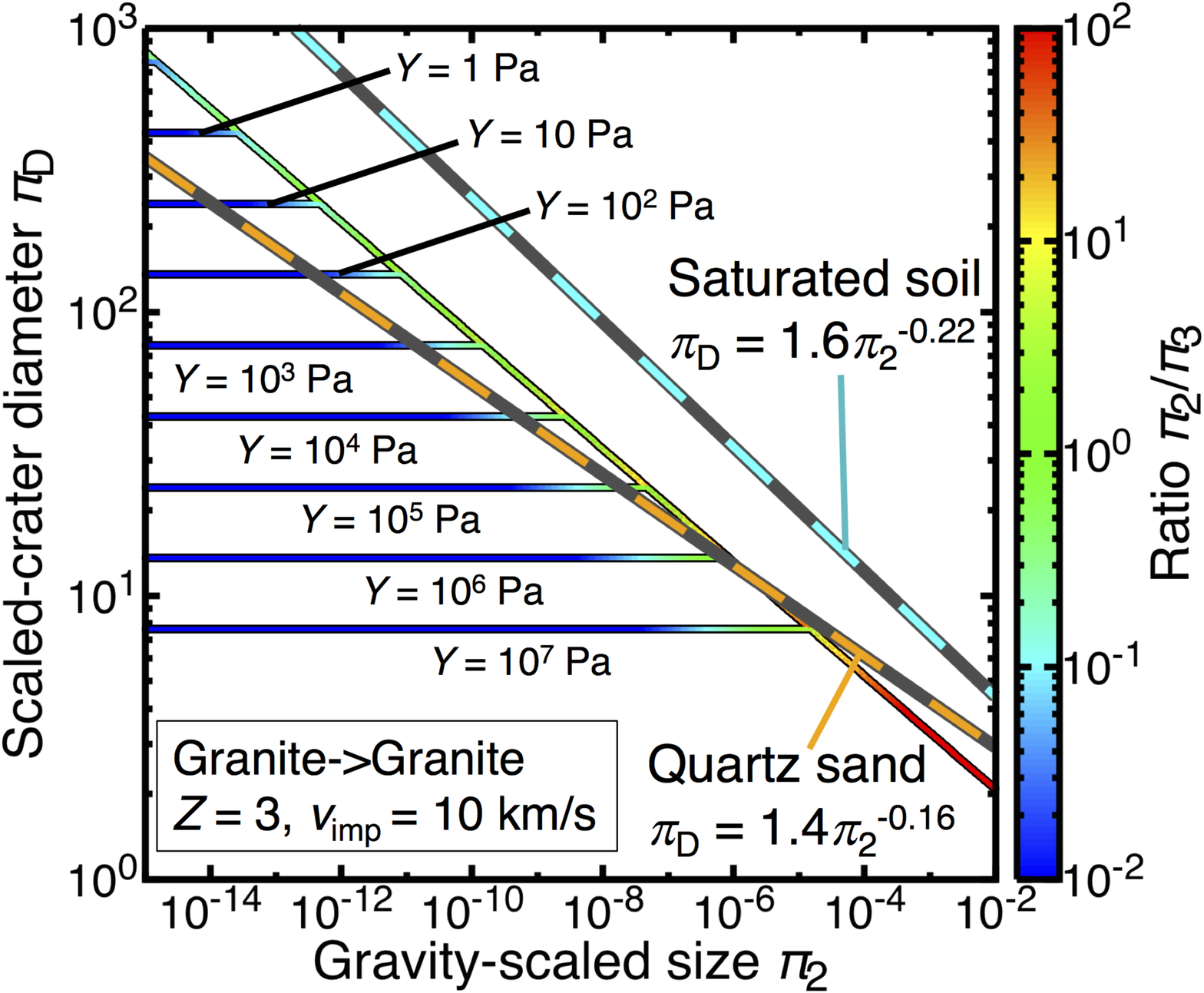}
	\end{center}
	\caption{As for Figs.\ \ref{fig:fig9} and \ref{fig:fig10}, but showing the effect of the strength $Y$ on the $\pi_D$ behavior versus $\pi_2$. 
	The scaled crater diameters are colored according to the ratio of $\pi_2$ to $\pi_3$.}
	\label{fig:fig11}
\end{figure}

Figures \ref{fig:fig9}--\ref{fig:fig11} show $R_{\rm tr}$ resulting from our model in the form of $\pi$-group scaling laws, along with the conventional results. 
A granite projectile and target were considered. 
We calculated the scaled crater diameter $\pi_D$ ($Y$ axis) and the gravity-scaled size $\pi_2$ ($X$ axis). 
We focused on the $Z$, $v_{\rm imp}$, and $Y$ dependences on $\pi_D$ in Figs.\ \ref{fig:fig9}--\ref{fig:fig11}, respectively. 
In Fig.\ \ref{fig:fig9}, $Z$ was varied from $2.01$ to $4$, $v_{\rm imp}$ was fixed at $10\ {\rm km\ s^{-1}}$, and no strength ($Y = 0$) was included. 
Although our prediction is sensitive to the exponent $Z$, the resulting $\pi_D$ values predicted by our model are contained within the two typical scaling lines pertaining to saturated soil and dry quartz sand \citep{Schmidt1987}, thereby strongly supporting the notion that our model accurately predicts transient crater radii. 
The differences in the materials for the conventional scaling laws correspond to the differences in $Z$ in our model. 
In Fig.\ \ref{fig:fig10}, $v_{\rm imp}$ was varied from $5$ to $100\ {\rm km\ s^{-1}}$, $Z$ was fixed at $3$, and no strength ($Y = 0$) was included. 
The different $v_{\rm imp}$ values yield different $\pi_D$ values within a factor of $1.6$ for the same value of $\pi_2$, suggesting that $\pi_2$ would not be a good measure to define the scale of impact events in terms of the $v_{\rm imp}$ dependence. 
The reason for this result is discussed in Section \ref{sec:4.4}. 
In Fig.\ \ref{fig:fig11}, $Y$ was varied from $1\ {\rm Pa}$ to $10^7\ {\rm Pa}$, $v_{\rm imp}$ was fixed at $10\ {\rm km\ s^{-1}}$, and $Z$ was fixed at $3$ to investigate the transition behavior from the gravity-dominated to the strength-dominated regime with decreasing $\pi_2$. 
In this calculation, we determined $R_{\rm tr}$ based on $\min(R_{\rm tr, grav}, R_{\rm tr, strength})$, where $R_{\rm tr, grav}$ and $R_{\rm tr, strength}$ are the transient crater radii calculated using Eqs.\ (\ref{eq:R_tr_kin_grav}) and (\ref{eq:R_tr_kin_strength}), respectively. 
The effect of material strength on crater formation becomes dominant when the ratio of $\pi_2$ to $\pi_3$ falls below $0.1$.


\section{Discussion}\label{sec:4}
\subsection{A verification via hydrocode modeling}\label{sec:4.1}
Here we discuss the limitations of our model. 
We assumed that the residual velocity is zero when $u_{\rm p, max}<u_{\rm p, th}$. 
Thus, the model predicts that no craters are produced for low-velocity impacts, regardless of the strength of the target material. 
This obviously contradicts the well-known fact that low-velocity impacts also produce impact craters when the target materials have relatively low strength. 
To address the limitations of the key assumption of the model that the residual velocity after a shock-release cycle is the origin of the excavation flow, we also perform a numerical simulation based on the iSALE shock physics code \citep{Amsden1980, Ivanov1997, Wunnemann2006}. 
The results are described in the Supplementary Information. 
We confirmed that our analytical result (thick black line in Fig.\ \ref{fig:fig4}) reproduces well the numerical results in the thermal-pressure-dominated range (red dashed line in Fig.\ \ref{fig:fig4}). 
The $u_{\rm p,res}$ in the cold-pressure-dominated regime (blue dashed line in Fig.\ \ref{fig:fig4}) is, however, considerably slower than the numerical results. 
This result indicates a difference in the physics underlying the formation of an excavation flow during low-velocity impacts compared with that during the hypervelocity impacts discussed in this study. 
Consequently, the accuracy of the model prediction is relatively high in the thermal-pressure-dominated regime, which roughly corresponds to the condition $v_{\rm imp} > 2u_{\rm p,sw}$ (typically $v_{\rm imp} > 5\ {\rm km\ s^{-1}}$), although it would predict somewhat slower ejection velocities and smaller transient craters in the cold-pressure-dominated regime. 
Our model is therefore suitable for predicting impact outcomes after hypervelocity impact events on Mars-sized or larger planets, their satellites, and after typical collisions between asteroids in the main-belt region.


\subsection{High-speed cut-off of the ejection velocity in the normal excavation process}\label{sec:4.2}
It is widely considered that impact excavation can be divided into three stages depending on ejection timing, location, velocity, and pressure: (1) jetting, (2) spallation, and (3) normal excavation \citep[e.g.,][]{Melosh1989, Kurosawa2018}. 
The transition behaviors from jetting to spallation to normal excavation have been summarized by \citet{Kurosawa2018}. 
In this study, we have discussed normal excavation. 
The ejecta velocity distribution owing to normal excavation is expressed as a power-law relation, as discussed in Section \ref{sec:3.3}. 
However, a high-speed cut-off $v_{\rm ejmax}$ is required because the total kinetic energy carried away by the ejecta becomes infinite without such a cut-off \citep{Housen2011}. 
A higher ejection velocity than the cut-off value can be achieved by jetting and/or spallation \citep[e.g.,][]{Melosh1986, Vickery1993, Johnson2014, Kurosawa_et_al2015, Kurosawa2018}. 
Our model, which was constructed based on the framework initially proposed by \citet{Melosh1985b}, clearly predicts the residual velocity in the isobaric core to be at the high-speed cut-off; i.e., $v_{\rm ejmax} = C u_{{\rm p}0}^m$. 
The cut-off is typically $4\%$--$20\%$ of $v_{\rm imp}$ (Fig.\ \ref{fig:fig4}).


\subsection{Correspondence between the proposed model and conventional scaling laws}\label{sec:4.3}
The ejecta velocity distribution and crater growth have been discussed in terms of the point-source theory \citep[e.g.,][]{Housen1983, Holsapple1993}. 
By comparing Eq.\ (\ref{eq:v_ej_R}) and the predictions of point-source theory, $Z\ $is expressed as a function of the velocity-scaling exponent $\mu$ as follows:
\begin{equation}
Z=\frac{2-\mu}{\mu}. \label{eq:Z_mu_1}
\end{equation}
Since the allowable $\mu\ $range spans from $1/3$ (momentum scaling) to $2/3$ (energy scaling) as discussed in Section \ref{sec:Introduction}, $Z$ is estimated to range from $2$ to $5$. 
This range is consistent with the value of the product $nm$ (Table \ref{fig:table}), and these might be reasonable values to describe a cratering flow field \citep[e.g.,][]{Croft1980, Melosh1989}. 
Note that the relationship between $\mu$ and $Z$ is frequently given by \citep[e.g.,][]{Housen1983}
\begin{equation}
Z=\frac{1}{\mu}. \label{eq:Z_mu_2}
\end{equation}
for the frequently used assumptions discussed in Section \ref{sec:2.1}; i.e., the time-dependent strength of the excavation flow $\alpha(t)={\rm Const.}$ 
The difference in the relationship between Eqs.\ (\ref{eq:Z_mu_1}) and (\ref{eq:Z_mu_2}) implies that our model implicitly assumes a time-dependent $\alpha(t)$, although we do not necessarily explicitly address the functional form of $\alpha(t)$. 
Since the shapes of the streamlines of the excavation flow depend only on the exponent $Z$ in the $Z$-model (Eq.\ (\ref{eq:r_R_theta})), we were able to formulate the equations related to the cratering processes without an explicit expression for $\alpha(t)$.
In situ observations of the growth of the crater radius $R_{\rm cr}(t)$ in a laboratory setting allows us to determine the exponents $\mu$ and $Z$ for each impact \citep[e.g.,][]{Yamamoto2009, Yamamoto2017}. 
The crater radius as a function of time is expressed as follows:
\begin{equation}
	R_{\rm cr}(t) \propto t^{e_{\rm x}},
\end{equation}
where $e_{\rm x}$ is an exponent determined from laboratory measurements \citep{Cintala1999}. 
By comparison with Eq.\ (\ref{eq:R_t_Z}), the exponent $Z$ is related to $e_{\rm x}$ as
\begin{equation}
	Z=\frac{2-3e_{\rm x}}{e_{\rm x}}.
\end{equation}
If we assume $Z$ to range from $2$ to $5$, as discussed above, $e_{\rm x}$ becomes $0.25$--$0.40$. 
The range of $e_{\rm x}$ is consistent with the values of $e_{\rm x}$ measured in laboratory experiments with dry sand targets \citep{Cintala1999, Yamamoto2017}.


\subsection{Modified measure of the gravity-scaled size}\label{sec:4.4}
Here, we discuss the effect of $v_{\rm imp}$ on $\pi_D$ and the gravity-scaled size $\pi_2$. 
First, we obtain the transient crater radius $R_{\rm tr}$ in the framework of the $\pi$-group scaling laws from Eqs.\ (\ref{eq:pi_D}), (\ref{eq:pi_2}), and (\ref{eq:pi_D_K1}) to directly compare with our model [Eqs.\ (\ref{eq:R_tr_kin_grav}) and (\ref{eq:R_tr_kin_grav2})]. 
In the case of collisions between the same materials (i.e., $\rho_{\rm p}=\rho_{\rm t}$ and $2u_{{\rm p}0} = v_{\rm imp}$), $R_{\rm tr}$ is rewritten as 
\begin{equation}
	R_{\rm tr}=k_{10} R_{\rm p}^{1-\beta} g^{-\beta} u_{{\rm p}0}^{2\beta}, \label{eq:R_beta}
\end{equation}
where $k_{10}$ is a dimensionless constant described in \ref{sec:A4} and
\begin{equation}
	\beta=\frac{\mu}{\mu+2}=\frac{1}{Z+2}.\label{eq:beta_mu_Z}
\end{equation}
Note that we used Eq.\ (\ref{eq:pi_D_K1}) to obtain Eq.\ (\ref{eq:beta_mu_Z}). 
Thus, the dependences of $R_{\rm p}$ and $g$ on $R_{\rm tr}$ in our model are consistent with the $\pi$-group scaling laws when Eq.\ (\ref{eq:beta_mu_Z}) is valid. 
In contrast, the $u_{{\rm p}0}$ dependence on $R_{\rm tr}$ is quite different, as shown in Eqs.\ (\ref{eq:R_tr_kin_grav}), (\ref{eq:R_tr_kin_grav2}), and (\ref{eq:R_beta}), which is expected to produce factor $1.6$ dispersion against the same $\pi_2$ values in our model shown in Fig.\ \ref{fig:fig9}.

Second, we discuss the origin of the difference in the $u_{{\rm p}0}$ dependence on $R_{\rm tr}$. 
Since the residual velocity ultimately originates from irreversible shock heating (i.e., an increase in entropy), as discussed in Section \ref{sec:2}, the conversion efficiency from the initial kinetic energy injected by the projectile $E_{\rm proj}$ to the total kinetic energy in the excavation flow $E_{\rm res}$ is expected to strongly depend on $v_{\rm imp}$. 
When $u_{{\rm p}0}$ is greater than $u_{\rm p,sw}$, $E_{\rm res}$ is divided into three terms in the same way as used in the derivation of $E_{\rm kin}$:
\begin{equation}
	E_{\rm res}=E_{{\rm res}1}+E_{{\rm res}2}+E_{{\rm res}3},
\end{equation}
where
\begin{eqnarray}
	E_{{\rm res}1} &=& k_{11} u_{{\rm p}0}^{2m_{\rm t}} R_{\rm p}^3,\\
	E_{{\rm res}2} &=& k_{12} u_{{\rm p}0}^{2m_{\rm t}} R_{\rm p}^{2nm_{\rm t}},\\
	E_{{\rm res}3} &=& k_{13} u_{{\rm p}0}^{2m_{\rm c}} R_{\rm p}^{2nm_{\rm c}}.
\end{eqnarray}
If $u_{{\rm p}0}<u_{\rm p, sw}$, $E_{\rm res}$ is expressed as
\begin{equation}
	E_{\rm res}=E_{{\rm res}1}+E_{{\rm res}2},
\end{equation}
where
\begin{equation}
	E_{{\rm res}1} = k_{14} u_{{\rm p}0}^{2m_{\rm c}} R_{\rm p}^3,
\end{equation}
and
\begin{equation}
	E_{{\rm res}2} = k_{15} u_{{\rm p}0}^{2m_{\rm c}} R_{\rm p}^{2nm_{\rm c}}.
\end{equation}
As in Section \ref{sec:3.2}, we only show the proportional relations with dimensional constants $k_{11}$--$k_{15}$. 
The explicit expressions of the coefficients are presented in \ref{sec:A4}. 
Since $u_{{\rm p}0}$ is linearly proportional to $v_{\rm imp}$ \citep[e.g.,][]{Melosh1989}, $E_{\rm res}$ is not expressed as a simple linear function of $E_{\rm proj} = M_{\rm p} v_{\rm imp}^2/2$. 
Consequently, $\pi_D$ is not fully scaled by the gravity-scaled size in our model (Fig.\ \ref{fig:fig10}). 
In light of our key assumption (i.e., that the excavation flow is driven by the residual velocity of the shocked materials) \citep{Melosh1985b}, we propose a modified expression of the gravity-scaled size as follows:
\begin{equation}
	\pi_{2{\rm mod}} = \frac{1}{16}\left(\frac{4\pi}{3}\right)^{\frac{4}{3}} \frac{\rho_{\rm p} g D_{\rm p}^4}{E_{\rm res}}
	= \frac{0.42\rho_{\rm p} g D_{\rm p}^4}{E_{\rm res}}.
\end{equation}
We simply used $E_{\rm res}$ here instead of $E_{\rm proj}$ in the original form of $\pi_2$ [Eq.\ (\ref{eq:pi_2})].


\subsection{Insights into future laboratory/numerical experiments}\label{sec:4.5}
The proposed model is one of the simplest methods to predict the crater size when Tillotson EOS parameters are available. 
The Tillotson parameters can be obtained if the shock Hugoniot parameters are available \citep[e.g.,][]{Melosh1989}. 
Since both laboratory and numerical impact experiments are expensive and time-consuming, the new method could serve as a quick-look tool pertaining to crater size and would significantly aid in the design of laboratory and numerical experiments. 
An advantage of the new model is that the mechanics of the impact cratering processes, which cannot be addressed by dimensional analysis, are considered. 
This allows us to predict the tendencies of the impact outcomes as a function of a range of variables, as discussed below.

The dimensionless parameters, including the internal friction $f$ and the porosity $\phi$, could in principle be incorporated into our model in a straightforward manner. 
The decay exponent $n$ is expected to become larger with increasing $f$ and $\phi$ \citep[e.g.,][]{Mitani2003, Wunnemann2006, Bierhaus2013, Nagaki2016}.
The exponent $m$ is also expected to change and depend on $\phi$,~because $\phi$ affects the degree of irreversible shock heating \citep[e.g.,][]{Ahrens1972, Wunnemann2008}. 
The effects of $f$ and $\phi$ on both crater size and the ejecta velocity distribution would appear as a change in $Z$ in our model, because $Z$ is possibly controlled by the product $nm$, as discussed in Section \ref{sec:3.1}. 
\citet{Prieur2017} examined the effects of $f$ and $\phi$ on the transient crater size based on a number of numerical experiments, and they presented empirical equations pertaining to $\mu$ as functions of $f$ and $\phi$. 
Similar numerical experiments, focusing on the effects of $f$ and $\phi$~on the exponents $n$, $m$, and $Z$ using a shock physics code, may be useful in obtaining a physical interpretation of the empirical equations.

Recent impact experiments performed by \citet{Yamamoto2017} suggest that the velocity-scaling exponent $\mu$ also depends on $v_{\rm imp}$. 
They concluded that a higher $v_{\rm imp}$ tends to lead to a lower $\mu$.
This experimental result is understandable if the assumption $Z\sim nm$ is correct, as explained below. 
Although the exponent $n$ was treated as a constant $n=1.87$ throughout this study, the shock decay exponent $n$ is weakly dependent on $v_{\rm imp}$. 
According to a series of hydrocode simulations \citep{Pierazzo1997}, a higher $v_{\rm imp}$ leads to a larger $n$, implying that a higher $v_{\rm imp}$ leads to a smaller $\mu$ because $\mu$ is related to $Z$ [Eq.\ (\ref{eq:Z_mu_1})]. 
This hypothesis is consistent with the experimental results of \citet{Yamamoto2017}. 
A larger $v_{\rm imp}$, however, also causes a smaller $m$ ($m_{\rm t} < m_{\rm c}$). 
Thus, a complex behavior of $\mu$ as a function of $v_{\rm imp}$ is expected in reality. 
Further discussion of the $v_{\rm imp}$ dependence of $\mu$is beyond the scope of this study. 
Numerical simulations would significantly contribute to solving this problem.


\section{Conclusion}\label{sec:conclusion}
We have proposed a model to predict impact outcomes by combining the Maxwell $Z$-model and the residual velocity. 
In this study, we omitted some physics behind the cratering processes, such as the gradual change in the velocity vectors of the excavating target materials after a shock-release cycle, the neglect of the effects of velocity difference between adjacent stream tubes, and metal-like targets with a constant yield strength $Y$. 
These simplifications allowed us to obtain analytical solutions.
The new model allows us to analytically calculate the ejecta velocity distribution, the time variation of crater radii, and transient crater radii for a given impact condition based on a set of input parameters, including the exponents $Z$, $n$, and $m$. 
By analogy with the Maxwell $Z$-model, we propose that the exponent $Z$, which controls the shapes of streamlines in the excavation flow, is related to the product of the exponents $n$ and $m$. 
Our model is combined with the widely used point-source theory through the relation between $Z$ and the velocity-scaling exponent $\mu$. 
The impact outcomes predicted by the new model seem to yield reasonable trends compared with previous results. 
The new analytical model could aid in the design of a future interactive study comparing laboratory and numerical experiments to obtain a better understanding of the controls on impact outcomes.

\section*{Acknowledgements}
We thank Hiroki Senshu for useful discussions. 
We also thank the developers of iSALE, including G.\ Collins, K.\ W\"{u}nnemann, B.\ Ivanov, J.\ Melosh, and D.\ Elbeshausen. 
We appreciate the suggestions by Boris Ivanov that helped us greatly improve the manuscript, and we thank Oded Aharonson for handling of this manuscript as the journal editor. 
We also acknowledge useful discussions at a workshop on planetary impacts held at Kobe University. 
KK is supported by JSPS KAKENHI Grant Nos.\ 17H01176, 17H02990, 17H01175, and 17K18812. 
ST is supported by JSPS KAKENHI Grant No.\ 16H06478.


\appendix


\section{Expressions of the thermal and cold pressures, $P_{\rm thermal}$ and $P_{\rm cold}$}\label{sec:A1}
We employed the Tillotson EOS to calculate the residual velocity, as discussed in Section \ref{sec:2.2}. 
The thermal and cold pressures, $P_{\rm thermal}$ and $P_{\rm cold}$, are given by \citep{Tillotson1962}
\begin{equation}
	P_{\rm thermal}(E,\rho)=\left[a+\frac{b}{\left(\frac{E}{E_0 \eta^2}+1\right)}\right]\rho E,
\end{equation}
and
\begin{equation}
	P_{\rm cold}(\rho)= A\mu + B\mu^2,
\end{equation}
where $\eta = \frac{\rho}{\rho_0}$, $\mu=\rho-1$, and $a$, $b$, $A$, $B$, and $E_0$ are the Tillotson parameters.


\section{Expression of $f(Z)$}\label{sec:A2}
From simple geometric considerations, the geometric factor $f(Z)$, which is the ratio of the total travel distance $L$ along a given streamline to the horizontal distance $R$ from the impact point, is expressed as
\begin{equation}
	f(Z) = \int_0^{\frac{\pi}{2}} \sqrt{ \left(\frac{\sin\theta}{Z-2}\right)^2 (1-\cos\theta)^{\frac{2(3-Z)}{Z-2}} + (1-\cos\theta)^{\frac{2}{Z-2}} }d\theta.
\end{equation}


\section{Definition of the threshold velocity, $u_{\rm p,th}$}\label{sec:A3}
We assumed that the residual velocity $u_{\rm p,res}$ is zero when the peak particle velocity $u_{\rm p,max}$ is slower than the threshold velocity $u_{\rm p,th}$. 
In other words, our model is only valid when the hypersonic condition applies, which corresponds to the condition that the wave speed $U$ is higher than the longitudinal sound speed $C_{\rm L}$. 
This definition is the same as that employed in a previous study \citep[][p.\ 38, fig.\ 3.7]{Melosh1989}]. 
The threshold velocity $u_{\rm p,th}$ is estimated by application of the widely used linear velocity relation \citep[e.g.,][]{Melosh1989}, $U = C_{\rm 0} + s u_{\rm p,max}$. 
Thus, $u_{\rm p, th}$ is approximated as
\begin{equation}
	u_{\rm p, th}=\frac{C_{\rm L} - C_0}{s}.
\end{equation}


\section{Explicit expressions of the coefficients}\label{sec:A4}
Here, we describe the coefficients from $k_1$ to $k_{15}$. 
The definitions of the variables are described in the main text.
\begin{eqnarray}
	k_1 &=& 2\pi \left(\frac{Z-2}{Z+1}\right)\rho_{\rm t},\\
	k_2 &=& \pi \left(\frac{Z-2}{Z+1}\right)\rho_{\rm t} C_{\rm t}^2,\\
	k_3 &=& \pi \left(\frac{Z-2}{2nm_{\rm t} - Z -1}\right)\rho_{\rm t} C_{\rm t}^2,\\
	k_4 &=& k_3 u_{\rm p, sw}^{\frac{2nm_{\rm t}-Z-1}{n}},\\
	k_5 &=& \pi \left(\frac{Z-2}{2nm_{\rm c} - Z -1}\right)\rho_{\rm t} C_{\rm c}^2 
		\left(u_{\rm p, sw}^{\frac{2nm_{\rm c}-Z-1}{n}} - u_{\rm p, th}^{\frac{2nm_{\rm c}-Z-1}{n}}\right),\\
	k_6 &=& \pi \left(\frac{Z-2}{Z+1}\right)\rho_{\rm t} C_{\rm c}^2,\\
	k_7 &=& \pi \left(\frac{Z-2}{2nm_{\rm c} - Z -1}\right)\rho_{\rm t} C_{\rm c}^2,\\
	k_8 &=& k_7 u_{\rm p, th}^{\frac{2nm_{\rm c}-Z-1}{n}},\\
	k_9 &=& \pi \left[\frac{Z^2-4Z+4}{Z(Z+2)}\right]\rho_{\rm t},
\end{eqnarray}
\begin{eqnarray}
	k_{10} &=& 4^\beta \left(\frac{4\pi}{3}\right)^{\frac{1-\beta}{3}}K_1,\\
	k_{11} &=& \frac{\pi}{3} \rho_{\rm t} C_{\rm t}^2,\\
	k_{12} &=& \frac{\pi}{2nm_{\rm t}-3} \rho_{\rm t} C_{\rm t}^2 
		\left( R_{\rm p}^{-(2nm_{\rm t}-3)} - R_{\rm sw}^{-(2nm_{\rm t}-3)} \right),\\
	k_{13} &=& \frac{\pi}{2nm_{\rm c}-3} \rho_{\rm t} C_{\rm c}^2 
		\left( R_{\rm sw}^{-(2nm_{\rm c}-3)} - R_{\rm th}^{-(2nm_{\rm c}-3)} \right),\\
	k_{14} &=& \frac{\pi}{3} \rho_{\rm t} C_{\rm c}^2,
\end{eqnarray}
and
\begin{equation}
	k_{15} = \frac{\pi}{2nm_{\rm c}-3} \rho_{\rm t} C_{\rm c}^2 
		\left( R_{\rm p}^{-(2nm_{\rm c}-3)} - R_{\rm th}^{-(2nm_{\rm c}-3)} \right),
\end{equation}
where
\begin{equation}
	R_{\rm sw} = R_{\rm p} \left(\frac{u_{{\rm p}0}}{u_{\rm p, sw}}\right)^{\frac{1}{n}},
\end{equation}
and
\begin{equation}
	R_{\rm th} = R_{\rm p} \left(\frac{u_{{\rm p}0}}{u_{\rm p, th}}\right)^{\frac{1}{n}}.
\end{equation}


\bibliographystyle{elsarticle-harv}




\end{document}